\documentclass[lettersize,journal]{IEEEtran}
\usepackage{amsmath,amsfonts}
\usepackage{amssymb}
\usepackage{algorithm}
\usepackage{array}
\usepackage{textcomp}
\usepackage{stfloats}
\usepackage{caption}
\usepackage{subfigure}
\usepackage{url}
\usepackage{bm}
\usepackage[noend]{algpseudocode}
\usepackage{amsfonts}
\usepackage{color}
\usepackage{diagbox}
\usepackage{multirow}
\usepackage{booktabs}
\usepackage{verbatim}
\usepackage{hyperref}
\usepackage{verbatim}
\usepackage{graphicx}
\usepackage{cite}
\usepackage{enumerate}

\allowdisplaybreaks[4]
\begin{document}
\bibliographystyle{IEEEtran}
\title{Double Low-Rank 4D Tensor Decomposition for Circular RIS-Aided mmWave MIMO-NOMA System Channel Estimation in Mobility Scenarios}

	\author{
	Wanyuan~Cai,
	~Xiaoping~Jin,
	~Youming~Li,
	~Menglei~Sheng,
	~Mingjun~Huang,
	~Qinke~Qi,
	~and~Qiang~Guo
%
%
}

\markboth{~Vol.~XX, No.~X, MONTH~2025}%
{Shell \MakeLowercase{\textit{et al.}}: A Sample Article Using IEEEtran.cls for IEEE Journals}


\maketitle

\begin{abstract}
Channel estimation is not only essential to highly reliable data transmission and massive device access but also an important component of the integrated sensing and communication (ISAC) in the sixth-generation (6G) mobile communication systems. In this paper, we consider a downlink channel estimation problem for circular reconfigurable intelligent surface (RIS)-aided millimeter-wave (mmWave) multiple-input multiple-output non-orthogonal multiple access (MIMO-NOMA) system in mobility scenarios. First, we propose a subframe partitioning scheme to facilitate the modeling of the received signal as a fourth-order tensor satisfying a canonical polyadic decomposition (CPD) form, thereby formulating the channel estimation problem as tensor decomposition and parameter extraction problems. Then, by exploiting both the global and local low-rank properties of the received signal, we propose a double low-rank 4D tensor decomposition model to decompose the received signal into four factor matrices, which is efficiently solved via alternating direction method of multipliers (ADMM). Subsequently, we propose a two-stage parameter estimation method based on the Jacobi-Anger expansion and the special structure of circular RIS to uniquely decouple the angle parameters. Furthermore, the time delay, Doppler shift, and channel gain parameters can also be estimated without ambiguities, and their estimation accuracy can be efficiently improved, especially at low signal-to-noise ratio (SNR). Finally, a concise closed-form expression for the Cramér-Rao bound (CRB) is derived as a performance benchmark. Numerical experiments are conducted to demonstrate the effectiveness of the proposed method compared with the other discussed methods.
\end{abstract}


\section{Introduction}
With the advancement of the Internet of Things (IoT) paradigm, demands for ultra-high data rate transmission, ultra-dense connection, and high precision positioning are increasingly intensifying, prompting academia and industry to seek solutions through sixth-generation (6G) mobile communications \cite{10054381} \cite{10580974}. Power-domain non-orthogonal multiple access (NOMA), one of the most promising next generation multiple access (NGMA) candidate technologies for 6G, enables serving multiple users in the same time-frequency resource block. When combined with millimeter-wave (mmWave) and multiple-input multiple-output (MIMO) technologies, it is expected to significantly enhance data transmission capabilities, connection density, and spectral efficiency of the 6G communication systems \cite{9693417}. However, the inherent weak penetration and large attenuation characteristics of mmWave make it susceptible to blockage along the line-of-sight (LoS) path, resulting in numerous communication blind spots \cite{9847080}. 

Fortunately, the reconfigurable intelligent surface (RIS) technology has emerged as a promising solution to overcome LoS path obstruction by generating additional signal propagation paths, thereby breaking up the performance bottleneck of the system \cite{10077119}. The RIS, an artificial structure with controllable electromagnetic properties, is typically comprised of passive reflecting unit cells, and can adaptively customize the electromagnetic responses and achieve desirable beamforming gain by adjusting propagation characteristic of the incident wave, such as amplitude, phase, and polarization \cite{10077119} \cite{10547354}  \cite{9206044}. Nevertheless, achieving optimal RIS control and obtaining desirable beamforming gain rely on accurate channel state information (CSI). Furthermore, the angle and time delay information contained in CSI can be used to realize high precision positioning for user equipment \cite{9540372} \cite{10772118}, which is the important component of the integrated sensing and communication (ISAC) of 6G mobile network. Therefore, how to achieve accurate channel estimation for the RIS-aided mmWave MIMO-NOMA system has become an essential problem.

Recently, various channel estimation methods have been proposed for RIS-aided mmWave MIMO systems. Owing to the limited scattering in the mmWave propagation environments, the channel inherently exhibits sparse characteristics \cite{9815098}. Consequently, the channel estimation problem  is formulated as a compressed sensing (CS) equation and solved by various methods, including orthogonal matching pursuit (OMP)-based method \cite{9354904}, sparse Bayesian learning (SBL)-based method \cite{10506751}, and approximate message passing-based method \cite{10360256}. To overcome the grid mismatch problem inherent in discrete CS-based methods, multi-dimensional atomic norm minimization (ANM)-based methods are proposed in \cite{9398559} and \cite{10145065} by formulating the problem as semidefinite programming (SDP) problems solved in the continuous domain. In view of the high computational complexity of the SDP problem, a reweighted partially decoupled atomic norm minimization-based method is proposed in \cite{10630591} to reduce the complexity while improving the accuracy.
	
Apart from sparsity, there is another notable property of the channel matrix, i.e. the low-rank characteristic. In \cite{9475488}, the nuclear norm is introduced to exploit the low-rank property of the RIS-aided mmWave MIMO channel, which achieves considerable estimation performance improvement. To better exploit the underlying low-rank structures and achieve simultaneous channel parameter decoupling, the tensor modeling framework, which has been successfully employed in various domains \cite{10517663} \cite{8894531},
is introduced to formulate the received signal as a high-dimensional tensor. In \cite{9361077} and \cite{10251466}, the received signal is modeled as a third-order tensor admitting parallel factor (PARAFAC) decomposition form. The decomposition is realized by an iterative alternating scheme. In \cite{10552118}, the received signal is formulated as a low-rank third-order tensor admitting canonical polyadic decomposition (CPD) form to divide the angles and time delay into corresponding factor matrices. Then, the alternating least squares (ALS) method is used for tensor decomposition, while the correlation-based method is used for channel parameter extraction. To address the sensitivity to numerical ill-conditioning problem existing in the normal equation of ALS, the CP-ALS-QR proposed in \cite{minster2023cp} can be applied to improve the decomposition accuracy of the CPD while enhancing stability.
Nevertheless, the performance of the iterative methods mentioned above suffers from the global optimality problem and severely relies on initialization \cite{9835123}. In \cite{10772118}, the subspace-based CPD method is proposed to resolve these problems by exploiting the Vandermonde structure behind the received signal and only uses basic linear algebra, avoiding initialization and iteration. However, these methods almost adopt passive RIS with regular topology structure, i.e. uniform linear/planar array (ULA/UPA), which can only achieve cascade channel or coupled channel angle estimation \cite{10413280}. Specifically, due to the lack of transceiver capability in RIS, channel parameters are mutually coupled while the RIS manifold cannot provide sufficient conditions for unique decoupling, causing ambiguity in channel information \cite{9398559}. Therefore, it cannot support further applications, such as localization or tracking \cite{10413280}.

To eliminate the ambiguity, a twin RIS with two cooperative RIS planes and a conformal RIS arranged nonlinearly on curved surfaces are proposed in \cite{9540372} and \cite{9774328}, respectively. However, these works either introduce extra hardware or rely on highly complex optimization algorithms. Accordingly, a circular RIS is proposed in \cite{10413280} by extending the circular antennas concepts into RIS designing \cite{1291780} \cite{1589932}, which can capture more wireless information by exploiting the physical and electromagnetic degrees of freedom. Then, a RIS training pattern is tailored based on the principle of phase mode excitation to assist channel parameter decoupling. Although the method in \cite{10413280} can achieve unique decoupling without introducing extra hardware, its RIS training pattern causes small inter spacing of RIS reflecting unit cells. Specifically, the number of training frames for the RIS training pattern increases rapidly with the radius of the circular RIS. When the radius is set larger, it not only decreases the parameter estimation performance but also results in a large dimension of the processed data, greatly increasing the computational complexity. Accordingly, in \cite{10413280}, $256$ RIS reflecting unit cells are arranged circularly with a radius of the wavelength of $28$ GHz carrier frequency. Such close spacing between reflecting unit cells not only causes severe mutual coupling \cite{10440504}, but also poses a challenge to manufacturing. Besides, the aforementioned methods are restricted to static scenarios, ignoring the Doppler effect in mobility scenarios.

Indeed, wireless communication in mobility scenarios is common, in which the faster movement velocity causes the larger Doppler frequency shift, leading to rapid channel changes in a short period of time \cite{9815098}. Thus, frequent channel estimation is required to mitigate the phase distortion caused by the Doppler shift, which will lead to overloaded pilot signaling overhead \cite{8410591}. Although some existing works consider this problem and use the state-of-the-art methods to achieve angles and Doppler shift estimation, such as the PARAFAC-based method \cite{9606606} and ALS-based method \cite{8693958}, they neglect the frequency selectivity fading caused by time delay. In \cite{9102449}, \cite{9445013}, and \cite{9815098}, the OMP-based, SBL-based and ALS-based methods considering angle of arrival (AoA), angle of departure (AoD), time delay, and Doppler shift are proposed to enhance the universality of the system. Nevertheless, these methods ignore the LoS path obstruction problem. In \cite{10153406}, a more general RIS-assisted mmWave system channel estimation problem comprehensively considering AoA, AoD, time delay, and Doppler shift is formulated and solved by an OMP-based method. However, it only achieves cascade channel estimation and can not achieve unique parameter decoupling. Overall, the methods aforementioned either do not completely considering the above parameters, introduce extra hardware, rely on highly complex methods, are unable to achieve unique decoupling, suffer from severe mutual coupling between RIS elements, or require further estimation performance improvement.

Motivated by these, this paper proposes a double low-rank 4D tensor decomposition-based channel estimation method for circular RIS-aided mmWave MIMO-NOMA system in mobility scenarios, which efficiently exploits both the global and local low-rank properties of the received signal to improve the channel estimation accuracy. By leveraging the features of circular RIS structure, we achieve unique decoupling of angle parameters, while making the number of training time slots independent of the radius of the circular RIS, avoiding severe mutual coupling and manufacturing difficulties caused by insufficient inter spacing of RIS reflecting unit cells. Furthermore, based on the free-space propagation feature, we achieve the delay, Doppler shift, and cascade gain parameters estimation without ambiguities. The main contributions of this paper are summarized as follows:

1) We establish a circular RIS-aided mmWave MIMO-NOMA
time-varying system, incorporating the Doppler shift caused by mobility scenarios and introducing the circular RIS and NOMA technology to solve the obstruction of LoS path while removing the estimation uncertainties induced by linear passive RIS arrays and improving the spectral efficiency, respectively. Moreover, we propose a subframe partitioning scheme without changing the standard frame structure, which further refines the subframe. 

2) Based on the subframe partitioning scheme, we formulate the received signal as a fourth-order tensor meeting with CPD form, which divides angles, time delay, and Doppler shift into corresponding factor matrices. Accordingly, we transform the channel parameters estimation as tensor decomposition and parameter extraction problems. It is worth mentioning that by leveraging the tensor modeling and the Jacobi-Anger expansion of the circular RIS, we make the number of training time slots independent of the radius of the circular RIS, overcoming the mutual coupling problem and manufacturing difficulties in \cite{10413280}.

3) We propose a double low-rank 4D tensor decomposition model, which can improve the decomposition accuracy, especially at low signal-to-noise ratio (SNR). Specifically, by exploiting both the global and local low-rank properties of the received signal, we can simultaneously achieve noise suppression and accurate factor matrices restoration. To improve the robustness to noise at low SNR, we model the tail of Gaussian noise as sparse noise, which is incorporated into
the proposed double low-rank 4D tensor decomposition model, to enhance the adaptability of the proposed method in different scenarios. Then, the alternating direction method of multipliers (ADMM) is applied to deal with the proposed model.

4) By leveraging the special structure of the circular RIS, a two-stage parameter estimation method based on the Jacobi-Anger expansion of the circular RIS is proposed to uniquely decouple the angle parameters, achieving a favorable trade-off between accuracy and complexity, which cannot be supported by traditional linear RIS topologies. Besides, we also achieve estimation of other parameters without ambiguities. Furthermore, we derive the Cramér-Rao bound (CRB) as a performance benchmark by vectorizing the received signal, leading to a more concise form compared with traditional derivation similar to \cite{7914672}. Numerical experiments are conducted to demonstrate the effectiveness of the proposed method.

\textit{Notations:} Throughout this paper, scalars are denoted by lowercase letters, e.g. $a$, vectors by boldface lowercase letters, e.g. $\bf a$, matrices by boldface uppercase letters, e.g. $\bf A$, and tensors by Euler script letters, e.g. $\boldsymbol{ {\cal A}}$.  $[{\bf a}]_{(i)}$, $[{\bf A}]_{(i_1,i_2)}$, and $[\boldsymbol{ {\cal A}}]_{(i_1,\cdots, i_N)}$ denote the $i$-th entry of vector $\bf a$, $(i_1,i_2)$-th entry of matrix $\bf A$, and $(i_1,\cdots, i_N)$-th entry of tensor $\boldsymbol{ {\cal A}}$, respectively. The operations $(\cdot)^T$, $(\cdot)^*$, $(\cdot)^H$, $(\cdot)^{-1}$, $(\cdot)^{\dagger}$, $\rm Tr(\cdot)$ denote the transpose, conjugate, Hermitian, inverse, pseudo-inverse, and trace, respectively. $\Vert \cdot \Vert_1$ and $\Vert \cdot \Vert_F$ denote $\ell_1$ norm and Frobenius norm, respectively. $\circ$, $\otimes$, $\odot$, $\times_n$, $\langle\cdot,\cdot \rangle$, and $\mathbb E(\cdot)$ represent the outer product, Kronecker product, Khatri-Rao product, $n$-mode product, inner product, and expectation operator, respectively.  ${\rm diag}({\bf a})$, ${\rm diag}_n({\bf A})$, ${\rm undiag}({\bf A})$, ${\rm vec}({\bf A})$, and ${\rm unvec}_{M\times N}(\bf a)$ denote the diagonal matrix formed by vector $\bf a$, diagonal matrix formed by the $n$-th row of matrix $\bf A$, vector comprised of the diagonal element of matrix $\bf A$, vector formed by stacking matrix $\bf A$ column by column, and $M \times N$ matrix formed by reshaping a $MN \times 1$ vector $\bf a$ column by column, respectively. ${\bf 0}_{M \times N}$ and ${\bf I}_N$ denote the $M \times N$ matrix with all zero entries and the $N \times N$ identity matrix, respectively. $j = \sqrt{-1}$ is an imaginary symbol.

\section{System Model}\label{Sec_System_Model}
\begin{figure}[htbp]
	\centering
	\vspace{-1em}
	\includegraphics[width=2.6 in]{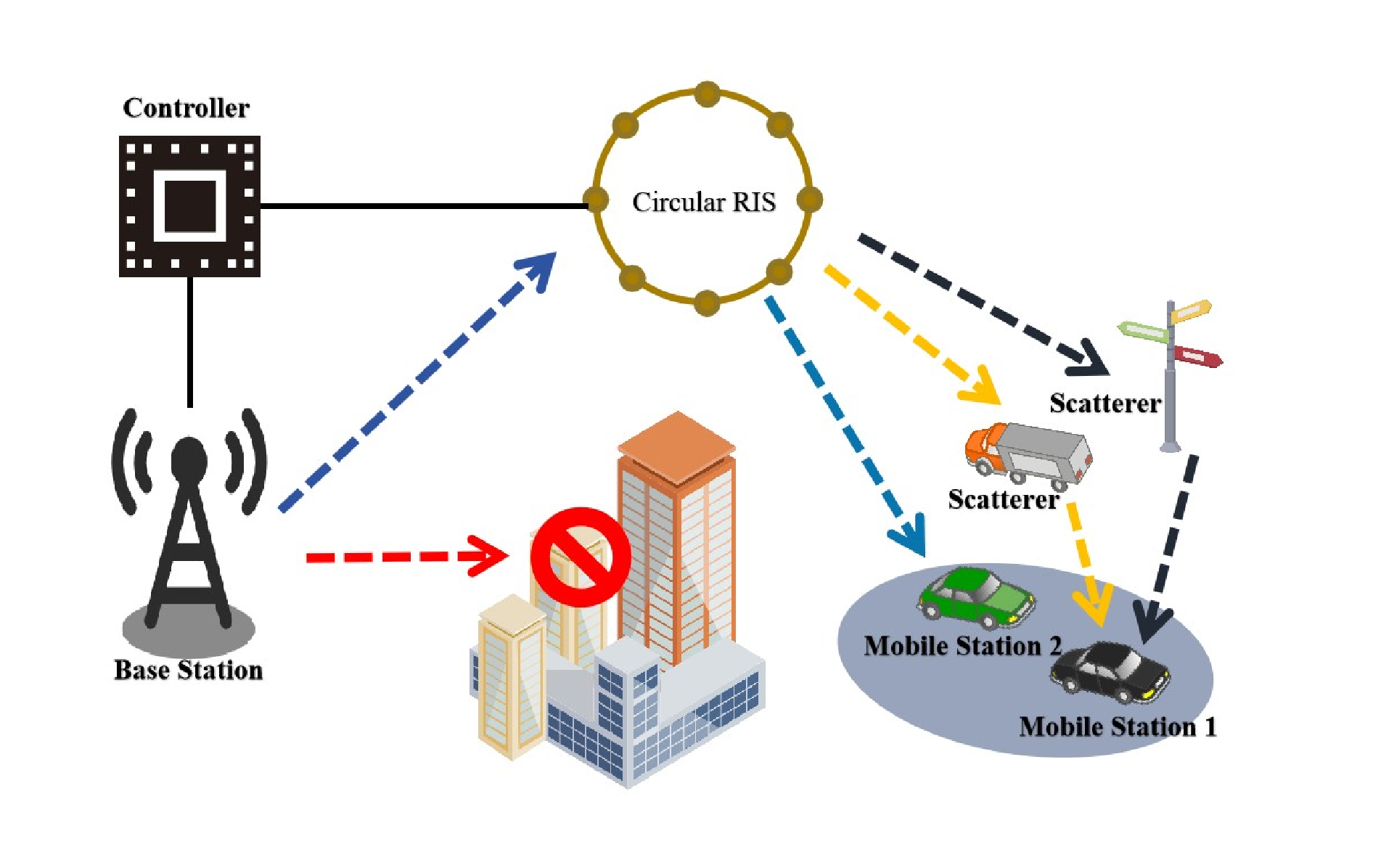}
	\caption{System model.}
	\label{System_pitcture}
	\vspace{-2.3em}
\end{figure}

We consider a downlink circular RIS-aided mmWave MIMO-NOMA system in a mobility scenario, as shown in Fig. \ref{System_pitcture}. The mobile stations (MSs) transmit large amounts of data at high transmission rates while moving, such as extensive real-time road and MS status information collected by diverse sensors. Due to the obstruction of buildings, the direct (i.e. LoS) path between base station (BS) and MSs are blocked. Thus, a circular RIS is introduced to provide extra paths for communication. To strike a balanced trade-off between system performance and hardware complexity, a hybrid analog-digital architecture is applied in both the BS and MSs to facilitate the simultaneous transmission of $N_s$ parallel data streams \cite{6847111} \cite{9049103}. The BS is equipped with $N_{\rm BS}$ antennas and $M_{\rm BS}$ radio-frequency (RF) chains, while each MS is equipped with $N_{\rm MS}$ antennas and $M_{\rm MS}$ RF chains, with $M_{\rm BS} < N_{\rm BS}$ and $M_{\rm MS} <N_{\rm MS}$. Both the antennas of BS and MSs are arranged in a half-wavelength spaced uniform linear array (ULA). The circular RIS is consisted of $N_{\rm R}$ reflecting unit cells which are uniformly arranged around a circle of radius $r$, where the azimuth and position of the $n$-th element can be denoted by $\zeta_n=2\pi(n-1)/N_{\rm R}$ and ${\bf p}_n=r[{\rm cos}\zeta_n,{\rm sin}\zeta_n]^T$, respectively. In the system, the total number of NOMA subcarriers with central carrier frequency $f_c$ and bandwidth $f_s$ is assumed to be $N$, among which the first $K$ subcarriers are pilot subcarriers used for channel estimation.
\subsection{Channel Modeling}
In practice, the BS and RIS are typically installed at higher locations, resulting in limited local scattering. Accordingly, the BS-RIS channel can be considered quasi-static, in which the LoS path has dominant power. Thus, the impulse response (CIR) of the BS-RIS channel can be represented as 
\begin{equation}
{\bf G}[\tau] = \alpha {\bf a}_{\rm R}(\theta_{\rm BR}) {\bf a}_{\rm B}^T(\phi_{\rm BR}) \delta(\tau - \tau_{\rm BR})
\label{G_tau.eqn}
\end{equation} 
where $\alpha$, $\theta_{\rm BR}$, $\phi_{\rm BR}$, and $\tau_{\rm BR}$ denote the complex channel gain, AoA of RIS, AoD of BS, and time delay, respectively. ${\bf a}_{\rm R}(\cdot)$ and ${\bf a}_{\rm B}(\cdot)$ represent the array response corresponding to the RIS and BS, respectively, expressed as
\begin{equation}
{\bf a}_{\rm R}(\cdot) = \frac{1}{\sqrt{N_{\rm R}}}[e^{j k_0 {\bf p}_1^T {\bf d}(\cdot)}, \cdots, e^{j k_0 {\bf p}_{N_{\rm R}}^T {\bf d}(\cdot)}]^T
\label{ar_theta.eqn}
\end{equation} 
\begin{equation}
{\bf a}_{\rm B}(\cdot) = \frac{1}{\sqrt{N_{\rm BS}}}[1,e^{j\pi {\rm cos}(\cdot)}, \cdots, e^{j \pi(N_{\rm BS}-1){\rm cos}(\cdot)}]^T
\label{ab_phi.eqn}
\end{equation} 
where $k_0=2\pi/ \lambda_c$ and ${\bf d}(\cdot) = [{\rm cos}(\cdot), {\rm sin}(\cdot)]^T$ denote the wave number of wavelength $\lambda_c$ and the unit direction vector, respectively. By performing Fourier transform on \eqref{G_tau.eqn}, the BS-RIS channel in the frequency domain corresponding to the $k$-th subcarrier can be expressed as
\begin{equation}
{\bf G}_k = \alpha e^{-j2 \pi \frac{k}{N} f_s \tau_{\rm BR}} {\bf a}_{\rm R}(\theta_{\rm BR}) {\bf a}_{\rm B}^T(\phi_{\rm BR}) 
\label{G_k.eqn}
\end{equation} 

For the RIS-MS channel, the mobility of the MS induces changes in the channel between the RIS and the MS due to the effect of Doppler frequency shift. To capture this effect, we employ a multi-path geometric mmWave channel model with $L$ paths \cite{7400949}, whose CIR is expressed as 
\begin{equation}
{\bf H}[t,\tau] = \sum_{l=1}^{L} \beta_l e^{j2 \pi f_d^l t} {\bf a}_{\rm M}(\theta_{\rm RM}^l) {\bf a}_{\rm R}^T(\phi_{\rm RM}^l) \delta(\tau - \tau_{\rm RM}^l)
\label{H_t_tau.eqn}
\end{equation} 
where $\beta_l$, $f_d^l$, $\theta_{\rm RM}^l$, $\phi_{\rm RM}^l$, and $\tau_{\rm RM}^l$ denote the complex channel gain, Doppler shift, AoA of MS, AoD of RIS, and time delay of the $l$-th path of RIS-MS channel, respectively. Here, $f_d^l = \frac{f_c v }{v_c}{\rm cos}(\theta_{v}^l)$, where $v$, $v_c$, and $\theta_{v}^l$ denote the velocity of MS, velocity of light, and AoA of MS relative to its moving direction in the $l$-th path, respectively. ${\bf a}_{\rm M}(\cdot)$ represents the array response corresponding to the MS, given by
\begin{equation}
{\bf a}_{\rm M}(\cdot) = \frac{1}{\sqrt{N_{\rm MS}}}[1,e^{j\pi {\rm cos}(\cdot)}, \cdots, e^{j \pi(N_{\rm MS}-1){\rm cos}(\cdot)}]^T
\label{am_theta.eqn}
\end{equation} 
\begin{figure}[h]
	\centering
	\vspace{-2em}
	\includegraphics[width=2.5 in]{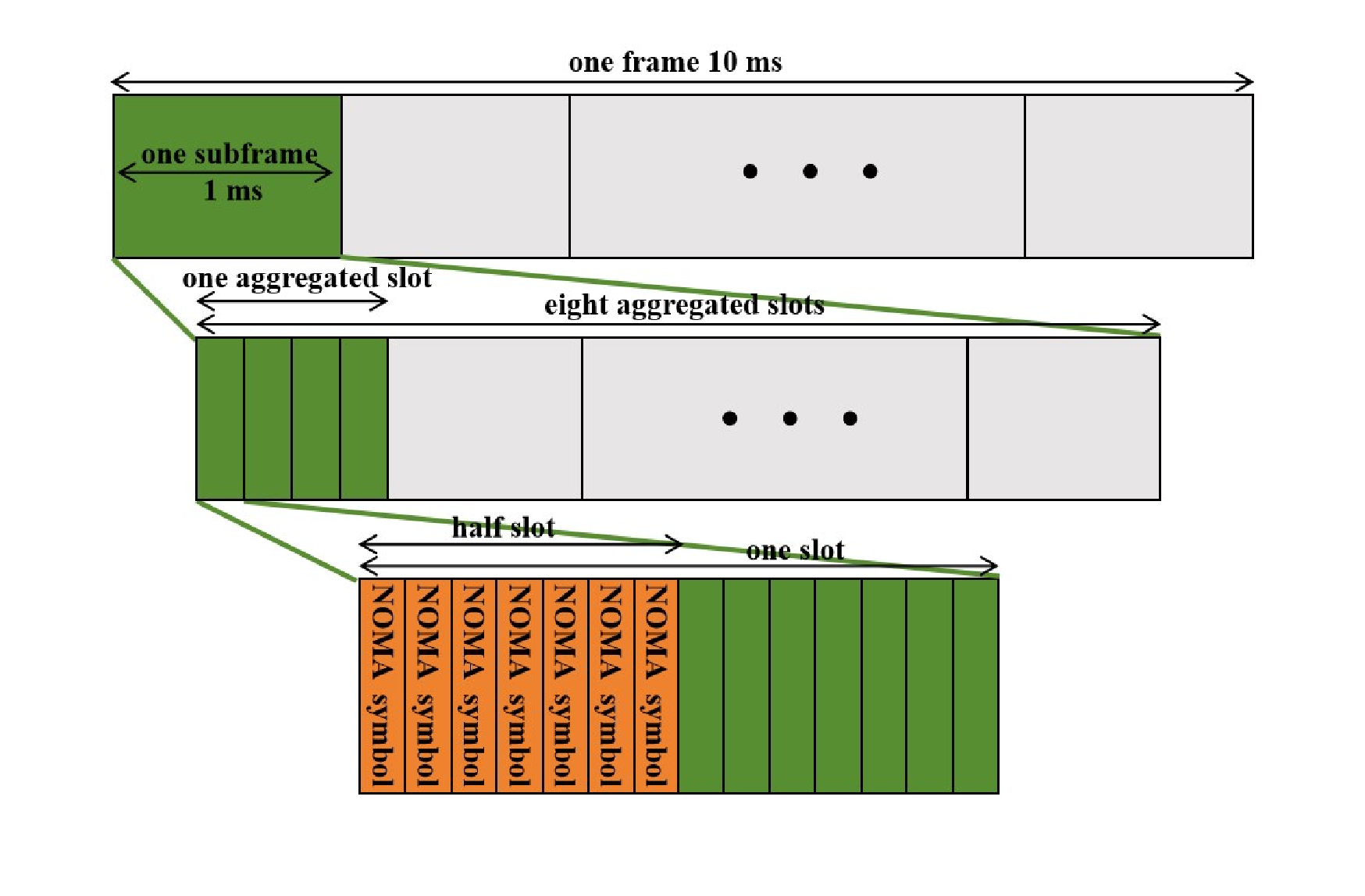}
	\caption{Subframe partitioning scheme.}
	\label{Frame_Structure}
	\vspace{-1.2em}
\end{figure}

In mobility scenarios, a faster moving velocity will introduce a higher Doppler frequency shift, causing rapid mmWave channel fluctuations within a short period of time. Owing to the numerology definition of the subcarrier spacing (SCS) and wide frequency band of the mmWave system, a larger SCS can be set to achieve shorter symbol duration time \cite{release18}. According to the technical specification in \cite{release18}, the SCS can be defined as $\Delta f = 480$ KHz, which means the duration of one slot (comprised of $14$ NOMA symbols) is $0.03125$ ms. Under this setting, the duration time of a wireless frame is $10$ ms, which contains $10$ subframes with each lasting $1$ ms. Then, based on the technical specification, a subframe partitioning scheme without changing the standard frame structure is proposed to refine the subframes, as shown in Fig. \ref{Frame_Structure}. For each refined subframe, it is comprised of $8$ aggregated slots, each of which is divided into $4$ slots, i.e. $8$ half-slots. Accordingly, the duration time of each aggregated slot is $0.125$ ms. When MS moves at a velocity of $90$ Km/h, the corresponding distance moved in one aggregated slot is only $0.3125$ cm, which can be absolutely ignored in far-field communication. Therefore, it is reasonable to assume that the channel gain, angles, time delay, and Doppler shift are unchanged during an aggregated slot, i.e. constant channel. It is worth mentioning that the SCS, the number of aggregated slots and half time slots in one frame can be appropriately adjusted according to the technical specification framework. Then, by denoting the number of aggregated slots in one subframe as $M$ and performing Fourier transform on \eqref{H_t_tau.eqn}, the frequency domain representation of high-mobility RIS-MS channel at the $k$-th subcarrier in the $m$-th aggregated slot is given by
\begin{equation}
{\bf H}_{k}[m] =  \sum_{l=1}^{L} \beta_l e^{-j2 \pi \frac{k}{N} f_s \tau_{\rm RM}^l} {\bf a}_{\rm M}(\theta_{\rm RM}^l) {\bf a}_{\rm R}^T(\phi_{\rm RM}^l) e^{j \omega_l m}
\label{HKM}
\end{equation} 
where $\omega_l = 2 \pi f_d^l T_s N_b N_{st}$, $T_s = 1 / \Delta f$, and $N_b$ and $N_{st}$ denote the number of NOMA symbols in one half slot and the number of half slots in one aggregated slot, respectively. 

\vspace{-1.5em}
\subsection{Signal Modeling}
In the NOMA system, the BS simultaneously transmits signals to multiple MSs at the same time using a shared time-frequency resource block. To improve the transmission rate, quadrature amplitude modulation (QAM) technology is incorporated into the system. Considering the $nst$-th half slot of the $m$-th aggregated slot, we denote the modulated data symbols residing on the $k$-th subcarrier of the $i$-th MS related to the $n_s$-th parallel data stream of $nb$-th NOMA symbol as $x_{i,n_s,nb,nst,k,m}$. Subsequently, the superposed signal $x_{n_s,nb,nst,k,m}$ can be obtained by
\begin{equation}
x_{n_s,nb,nst,k,m} = \sum_{i=1}^{N_{\rm U}} \sqrt{\iota_i P_t}x_{i,n_s,nb,nst,k,m} 
\label{superposed_signal.eqn}
\end{equation} 
where $N_{\rm U}$ denotes the total number of MSs using a shared time-frequency resource block, $P_t$ and $\iota_i$ denote the total power and the fraction of the total power assigned to the $i$-th MS, respectively. Accordingly, the received signal  ${\bf y}_{nb, nst,k,m}$ at $k$-th subcarrier of the $nb$-th NOMA symbol in the $nst$-th half slot of the $m$-th aggregated slot can be expressed as
\begin{equation}
\begin{aligned}
{\bf y}_{nb, nst,k,m} =& {\bf W}^T {\bf H}_k[m] {\rm diag}({\boldsymbol{\Psi}_{nst}}) {\bf G}_k {\bf F} {\bf x}_{nb,nst,k,m} \\
&+{\bf W}^T{\bf n}_{nb,nst,k,m} \in \mathbb{C}^{N_s \times 1}
\end{aligned}
\label{recieved_signal.eqn}
\end{equation} 
where ${\bf F} = {\bf F}_{\rm RF} {\bf F}_{\rm BB} \in {\mathbb{C}}^{N_{\rm BS} \times N_{s}}$ denotes the hybrid precoding matrix of BS comprised of analog RF precoder matrix ${\bf F}_{\rm RF}$ and digital baseband precoder matrix ${\bf F}_{\rm BB}$, ${\bf W} = {\bf W}_{\rm RF} {\bf W}_{\rm BB} \in {\mathbb{C}}^{N_{\rm MS} \times N_{s}}$ denotes the hybrid combining matrix of MS comprised of analog RF combining matrix ${\bf W}_{\rm RF}$ and digital baseband combining matrix ${\bf W}_{\rm BB}$, ${\bf x}_{nb,nst,k,m} = [x_{1,nb,nst,k,m},\cdots,x_{N_s,nb,nst,k,m}]^T \in \mathbb{C}^{N_s \times 1}$ denotes the parallel data streams transmitted at the $k$-th subcarrier of the $nb$-th NOMA symbol in the $nst$-th half slot of the $m$-th aggregated slot, $\boldsymbol{\Psi}_{nst} = [e^{j \xi_1},e^{j \xi_2},\cdots,e^{j \xi_{N_{\rm R}}}]^T$, $\xi_i \in [0,2\pi]$ denotes the phase shift vector of RIS at the $nst$-th half slot, and ${\bf n}_{nb,nst,k,m} \in \mathbb{C}^{N_{\rm MS} \times 1} \sim {\cal CN}(0,\sigma^2 {\bf I}_{N_{\rm MS}})$ represents the additive independent and identically distribution (i.i.d.) zero-mean circularly symmetric complex white Gaussian noise with variance $\sigma^2$. We assume that $\bf W$ and $\bf F$ remain constant within one subframe. 

Vectorizing \eqref{recieved_signal.eqn}, we can obtain 
\begin{equation}
\begin{aligned}
&{\bf y}_{nb, nst,k,m}  =  {\rm vec}({\bf y}_{nb, nst,k,m}) = (({\bf F} {\bf x}_{nb,nst,k,m}) \otimes {\bf W})^T\\
&~~~~~~~~ ({\bf G}_k^T \odot {\bf H}_k[m]) \boldsymbol{\Psi}_{nst} +{\bf W}^T{\bf n}_{nb,nst,k,m} \in \mathbb{C}^{N_s \times 1}
\end{aligned}
\label{vec_recieved_signal.eqn}
\end{equation} 
where ${\rm vec}(\cdot)$ denotes the vectorization operation, and ${\bf G}_k^T \odot {\bf H}_k[m]$ represents the cascade channel, which can be rewritten by substituting \eqref{G_k.eqn} and \eqref{HKM} as 
\vspace{-0.3em}
\begin{equation}
\begin{aligned}
&{{\bf H}}^{\bf G}_k[m] = {\bf G}_k^T \odot {\bf H}_k[m] \\ 
&= \sum_{l=1}^{L} \alpha \beta_l e^{-j2 \pi \frac{k}{N} f_s (\tau_{\rm RM}^l + \tau_{\rm BR})}  ({\bf a}_{\rm B}(\phi_{\rm BR}) \otimes {\bf a}_{\rm M}(\theta_{\rm RM}^l) )\\
&~~~~({\bf a}_{\rm R}^T(\theta_{\rm BR}) \odot {\bf a}_{\rm R}^T(\phi_{\rm RM}^l)) e^{j \omega_l m} \\
&= \sum_{l=1}^{L} \rho_l e^{-j2 \pi \frac{k}{N} f_s \tau_l}  {\bf a}_{s}(\phi_{\rm BR},\theta_{\rm RM}^l) {\bf a}_{r}^T(\theta_{\rm BR},\phi_{\rm RM}^l) e^{j \omega_l m}
\end{aligned}
\label{cascade_channel.eqn}
\vspace{-0.2em}
\end{equation} 
where $\rho_l = \alpha \beta_l$, $\tau_l = \tau_{\rm RM}^l + \tau_{\rm BR}$, ${\bf a}_{s}(\phi_{\rm BR},\theta_{\rm RM}^l) = {\bf a}_{\rm B}(\phi_{\rm BR}) \otimes {\bf a}_{\rm M}(\theta_{\rm RM}^l)$, and ${\bf a}_{r}(\theta_{\rm BR},\phi_{\rm RM}^l) = ({\bf a}_{\rm R}^T(\theta_{\rm BR}) \odot {\bf a}_{\rm R}^T(\phi_{\rm RM}^l))^T$ which can be rewritten as 
\begin{equation}
\begin{aligned}
&{\bf a}_{r}(\theta_{\rm BR},\phi_{\rm RM}^l) = ({\bf a}_{\rm R}^T(\theta_{\rm BR}) \odot {\bf a}_{\rm R}^T(\phi_{\rm RM}^l))^T\\
&=  \frac{1}{{N_{\rm R}}}[e^{j k_0 {\bf p}_1^T ({\bf d}(\theta_{\rm BR})+{\bf d}(\phi_{\rm RM}^l))},\cdots,e^{j k_0 {\bf p}_{N_{\rm R}}^T ({\bf d}(\theta_{\rm BR})+{\bf d}(\phi_{\rm RM}^l))}]^T\\
&=\frac{1}{{N_{\rm R}}}[e^{j2k_0 r{\rm cos}(\frac{\theta_{\rm BR}-\phi_{\rm RM}^l}{2}){\rm cos}(\frac{\theta_{\rm BR}+\phi_{\rm RM}^l}{2}-\zeta_1)},\cdots,\\
&~~~~~~~~~~e^{j2k_0 r{\rm cos}(\frac{\theta_{\rm BR}-\phi_{\rm RM}^l}{2}){\rm cos}(\frac{\theta_{\rm BR}+\phi_{\rm RM}^l}{2}-\zeta_{N_{\rm R}})}]^T
\end{aligned}
\label{ar_theta_phi.eqn}
\vspace{-0.2em}
\end{equation} 

Denoting $\theta_{eq}^l = \frac{\theta_{\rm BR}-\phi_{\rm RM}^l}{2}$ and $\phi_{eq}^l = \frac{\theta_{\rm BR}+\phi_{\rm RM}^l}{2}$, we have
\begin{equation}
\begin{aligned}
&{\bf a}_{r}(\theta_{eq}^l,\phi_{eq}^l) ={\bf a}_{r}(\theta_{\rm BR},\phi_{\rm RM}^l)\\
&= \frac{1}{{N_{\rm R}}}[e^{j2k_0 r{\rm cos} \theta^l_{eq} {\rm cos}(\phi_{eq}^l-\zeta_1)},\cdots,e^{j2k_0 r{\rm cos} \theta^l_{eq} {\rm cos}(\phi_{eq}^l-\zeta_{N_{\rm R}})}]^T
\end{aligned}
\label{ar_theta_phi_rewrite.eqn}
\end{equation} 

Thanks to the special structure of circular RIS, ${\bf a}_{r}(\theta_{\rm BR},\phi_{\rm RM}^l)$ in  \eqref{ar_theta_phi_rewrite.eqn} can be expanded into the sum of infinite terms of exponentials of trigonometric functions by Jacobi-Anger expansion, expressed as \cite{abramowitz1968handbook} \cite{8812953}
\vspace{-0.6em}
\begin{equation}
\begin{aligned}
\left[{\bf a}_{r}(\theta_{eq}^l,\phi_{eq}^l)\right]_{(n)} = \sum_{i=-\infty}^{+\infty} {j}^i {\rm J}_i(2k_0r{\rm cos}\theta^l_{eq})e^{-j\zeta_n i} e^{j\phi_{eq}^l i}
\end{aligned}
\label{jacobi-anger_infinit}
\end{equation} 
where ${\rm J}_i$ denote the $i$-th order Bessel function of the first kind. 

According to \cite{8812953}, the infinite series expansion can be approximated with considerable precision by summing finite terms, i.e. 
\vspace{-0.6em}
\begin{equation}
\begin{aligned}
\left[{\bf a}_{r}(\theta_{eq}^l,\phi_{eq}^l)\right]_{(n)} \approx  \sum_{i=-I}^{+I} {j}^i {\rm J}_i(2k_0r{\rm cos}\theta^l_{eq})e^{-j\zeta_n i} e^{j\phi_{eq}^l i}
\end{aligned}
\label{jacobi-anger_finit}
\vspace{-0.5em}
\end{equation} 
where $I> \frac{2\pi r}{\lambda}$. To achieve higher precision, we set $I = 2 \cdot {\rm ceil}(\frac{2\pi r}{\lambda})$ in this paper, where ${\rm ceil}(\cdot)$ denotes the upward rounding operator. Then, \eqref{jacobi-anger_finit} can be rewritten as 
\vspace{-0.3em}
\begin{equation}
\begin{aligned}
{\bf a}_{r} ( \theta_{eq}^l, \phi_{eq}^l) \approx  \boldsymbol{\Theta} {\bf J}_d(\theta_{eq}^l) {\bf v}(\phi_{eq}^l)
\end{aligned}
\label{jacobi-anger_matrix}
\end{equation} 
where the entries of $\boldsymbol{\Theta}$, ${\bf J}_d(\theta_{eq}^l)$, and ${\bf v}(\phi_{eq}^l)$ are given by 
\vspace{-0.3em}
\begin{equation}
\begin{aligned}
&\left[\boldsymbol{\Theta}\right]_{(n,i)} = j^i e^{-j \zeta_n i}, n = 1,\cdots N_{\rm R}, i = -I,\cdots,I \\
&\left[{\bf J}_d(\theta_{eq}^l)\right]_{(i,i)}= {\rm J}_i(2k_0r{\rm cos}\theta_{eq}^l), i = -I,\cdots,I\\
&\left[{\bf v}(\phi_{eq}^l)\right]_{(i)} = e^{j\phi_{eq}^l i},i = -I,\cdots,I
\end{aligned}
\label{jacobi-anger_matrix_entry}
\end{equation} 

Accordingly, parameters $\theta_{eq}^l$ and $\phi_{eq}^l$ are decoupled into a diagonal matrix only related to $\theta_{eq}^l$ and a column vector only related to $\phi_{eq}^l$, respectively, which will be used for parameter estimation later on.

\vspace{-1.3em}
\subsection{Tensor-Based Representation}
\vspace{-0.2em}
The received signal \eqref{vec_recieved_signal.eqn} encompasses various independent variables, i,e, angles, time delay, and Doppler shift variables, posing difficulties for parameter estimation using conventional methods. Thus, we formulate the received signals within one subframe as a fourth-order tensor satisfying CPD form, enabling the separation of variables into factor matrices and greatly enhancing the accuracy of parameter estimation. Specifically, by collecting the transmitted NOMA symbol at the pilot subcarriers in a half slot and defining ${\bf X}_{nst,k,m} = [{\bf x}_{1,nst,k,m},\cdots,{\bf x}_{N_b,nst,k,m}] \in \mathbb{C}^{N_s \times N_b}$, we can obtain
\begin{equation}
\begin{aligned}
&{\bf y}_{nst,k,m}  =  (({\bf F} {\bf X}_{nst,k,m}) \otimes {\bf W})^T({\bf G}_k^T \odot {\bf H}_k[m]) \boldsymbol{\Psi}_{nst}\\
&+ {\rm vec}({\bf W}^T{\bf N}_{nst,k,m} ) \in \mathbb{C}^{ N_s N_b \times 1},  k = 1,\cdots,K
\end{aligned}
\label{vec_recieved_signal_collecting.eqn}
\end{equation} 
where ${\bf y}_{nst,k,m} = [{\bf y}^T_{1,nst,k,m},\cdots,{\bf y}^T_{N_b,nst,k,m}]^T$ and ${\bf N}_{nst,k,m} = [{\bf n}_{1,nst,k,m},\cdots, {\bf n}_{N_b,nst,k,m}] \in \mathbb{C}^{N_{\rm MS} \times N_b}$. Note that the $N_s$ parallel data streams transmitted at pilot subcarriers remain independent within a half slot, but are identical in different half slots. Thus, we define ${\bf X}  = {\bf X}_{nst,k,m}$ and $\boldsymbol{\Upsilon} = ({\bf FX}) \otimes {\bf W}$. Then, by collecting all the phase shift vectors of RIS within one aggregated slot and defining the phase shift matrix $\boldsymbol{\Xi} = [\boldsymbol{\Psi}_{1},\cdots,\boldsymbol{\Psi}_{N_{st}}]$, the received signal at the $k$-th pilot subcarrier in the $m$-th aggregated slot is obtained by
\begin{equation}
\begin{aligned}
{\bf Y}_{k,m} & =  \boldsymbol{\Upsilon}^T({\bf G}_k^T \odot {\bf H}_k[m]) \boldsymbol{\Xi}\\
&~~~~ +{\rm Vec}^1_2(\boldsymbol{{\cal N}}_{k,m} \times_1 {\bf W}^T) \in \mathbb{C}^{N_s N_b \times N_{st}}
\end{aligned}
\label{matrix_recieved_signal_collecting.eqn}
\end{equation} 
where ${\bf Y}_{k,m} = [{\bf y}_{1,k,m}, \cdots, {\bf y}_{N_{st},k,m}]$, ${\rm Vec}^1_2(\cdot)$ denotes the vectorization operation for all frontal slices of a tensor, e.g. for a third-order tensor $\boldsymbol{{\cal A}} \in \mathbb{C}^{I_1 \times I_2 \times I_3}$, ${\rm Vec}^1_2(\boldsymbol{{\cal A}}) = [{\rm vec}(\boldsymbol{{\cal A}}(:,:,1)),\cdots, [{\rm vec}(\boldsymbol{{\cal A}}(:,:,I_3))] \in \mathbb{C}^{I_1 I_2 \times I_3}$, and $\boldsymbol{{\cal N}}_{k,m} \in \mathbb{C}^{N_{\rm MS} \times N_b \times N_{st}}$. Note that $\boldsymbol{\Xi}$ is the same in different aggregated slots. Substituting \eqref{cascade_channel.eqn} into \eqref{matrix_recieved_signal_collecting.eqn} and stacking the received signals of the first $K$ subcarriers in the $m$-th aggregated slot, we can obtain
\vspace{-0.3em}
\begin{equation}
\begin{aligned}
\boldsymbol{{\cal Y}}_{m} & = \sum_{l=1}^{L} \rho_l e^{j \omega_l m}  (\boldsymbol{\Upsilon}^T {\bf a}_{s}(\phi_{\rm BR},\theta_{\rm RM}^l)) \circ (\boldsymbol{\Xi}^T{\bf a}_{r}(\theta_{\rm BR},\phi_{\rm RM}^l)) \\  
&~~~~ \circ {\bf g}({\tau_l})  +{\rm Vec}^1_2(\boldsymbol{{\cal N}}_{m} \times_1 {\bf W}^T) \in \mathbb{C}^{N_s N_b \times N_{st}\times K}
\end{aligned}
\label{tensor_recieved_signal_collecting.eqn}
\end{equation} 
where ${\bf g}({\tau_l}) = [e^{-j2 \pi \frac{1}{N} f_s \tau_l},\cdots, e^{-j2 \pi \frac{K}{N} f_s \tau_l}]^T \in \mathbb{C}^{K \times 1}$ and $\boldsymbol{{\cal N}}_{m} \in \mathbb{C}^{N_{\rm MS} \times N_b \times N_{st} \times K}$. By defining ${\bf A} = [\boldsymbol{\Upsilon}^T {\bf a}_{s}(\phi_{\rm BR},\theta_{\rm RM}^1),\cdots, \boldsymbol{\Upsilon}^T {\bf a}_{s}(\phi_{\rm BR},\theta_{\rm RM}^L)]=[{\bf a}_1,\cdots,{\bf a}_L] \in \mathbb{C}^{N_s N_b \times L}$, ${\bf B} = [\boldsymbol{\Xi}^T{\bf a}_{r}(\theta_{\rm BR},\phi_{\rm RM}^1),\cdots,\boldsymbol{\Xi}^T{\bf a}_{r}(\theta_{\rm BR},\phi_{\rm RM}^L)] = [{\bf b}_1,\cdots, {\bf b}_L] \in \mathbb{C}^{N_{st} \times L}$, ${\bf C} = [\rho_1 {\bf g}({\tau_1}), \cdots, \rho_L {\bf g}({\tau_L})] =  [{\bf c}_1, \cdots, {\bf c}_L]\in \mathbb{C}^{K \times L}$ and stacking all the aggregated slots, we can formulate received signal $\boldsymbol{{\cal Y}} \in \mathbb{C}^{N_s N_b \times N_{st}\times K\times M} $ as a low-rank fourth-order tensor, whose CP decomposition form is expressed as 
\vspace{-0.3em}
\begin{equation}
\begin{aligned}
\boldsymbol{{\cal Y}} & = \sum_{l=1}^{L} {\bf a}_l \circ {\bf b}_l \circ {\bf c}_l \circ {\bf d}_l +{\rm Vec}^1_2(\boldsymbol{{\cal N}} \times_1 {\bf W}^T) 
\end{aligned}
\label{tensor_recieved_signal_four_order.eqn}
\end{equation} 
where ${\bf d}_l = [e^{j \omega_l},\cdots, e^{j \omega_l M}]^T \in \mathbb{C}^{M \times 1}$, and $\boldsymbol{{\cal N}} \in \mathbb{C}^{N_{\rm MS} \times N_b \times N_{st} \times K \times M}$. By defining ${{\bf D} = [{\bf d_1},\cdots,{\bf d}_L]}$ and $\boldsymbol{{\cal N}}^{\bf W} = {\rm Vec}^1_2(\boldsymbol{{\cal N}} \times_1 {\bf W}^T)  $, the received signal in \eqref{tensor_recieved_signal_four_order.eqn} can also be written as 
\begin{equation}
\begin{aligned}
\boldsymbol{{\cal Y}} &= \boldsymbol{{\cal I}} \times_1 {\bf A} \times_2 {\bf B} \times_3 {\bf C} \times_4 {\bf D} +\boldsymbol{{\cal N}}^{\bf W}\\
&= [\![{\bf A}, {\bf B},{\bf C}, {\bf D}]\!] + \boldsymbol{{\cal N}}^{\bf W} = \boldsymbol{{\cal Z}} + \boldsymbol{{\cal N}}^{\bf W}
\end{aligned}
\label{tensor_recieved_signal_modenpruduct.eqn}
\end{equation} 
where $\boldsymbol{{\cal Z}}$ denotes the received signal without noise interference. A detailed description of notations and foundations of tensor can be found in \cite{kolda2009tensor}.

\vspace{-0.5em}
\section{Double Low-Rank 4D Tensor Decomposition-Based Channel Estimation}
In this section, we introduce the proposed double low-rank 4D tensor decomposition-based channel estimation algorithm for the time-varying circular RIS-aided MIMO-NOMA system. First, we propose a double low-rank 4D tensor decomposition method to restore four factor matrices from the noisy received signal. Then, the channel parameters are estimated from the restored factor matrices. To reduce the complexity of high-precision two-dimensional search, a two-stage channel parameters estimation method is used.

\vspace{-1em}
\subsection{Double Low-Rank 4D Tensor Decomposition}
Factor matrices restoration is the key to channel parameters estimation, whose restoration accuracy directly affects the parameters estimation performance. Thus, we propose a double low-rank 4D tensor decomposition (DLR4DTD) method to enhance the factor matrices restoration performance by leveraging both the global and local low-rank properties of the received signal. Specifically, the Tucker decomposition is applied to exploit the global correlation among the received signal, which has superior noise suppression performance. Besides, the structured CP tensor decomposition is applied in the proposed model to characterize the local low-rank property of the mode-$1$ unfolding of the received signal, which simultaneously exploits the Vandermonde structure of the factor matrix to enhance the factor matrices restoration performance. Due to the normal distribution of Gaussian noise, the tail of the noise exhibits strong intensity and sparsity. Thus, to further improve the robustness to noise at low SNR, we model the tail of Gaussian noise as sparse noise, which can enhance the adaptability of the proposed method in different scenarios. The double low-rank 4D tensor decomposition model is shown as 
\begin{equation} 
\begin{aligned}
&\min_{\boldsymbol{{\cal Z}},\boldsymbol{{\cal S}}}{\Vert \boldsymbol{{\cal Y}} - \boldsymbol{{\cal Z}} -\boldsymbol{{\cal S}}\Vert}_F^2+\mu_1{\Vert \boldsymbol{{\cal Z}} \Vert}_{\rm SCPD}+\mu_2{\Vert \boldsymbol{{\cal S}} \Vert}_1 \\
&s.t. \quad \boldsymbol{{\cal Z}}=\boldsymbol{{\cal G}} \times_1 {\bf U}_1 \times_2 {\bf U}_2 \times_3 {\bf U}_3 \times_4 {\bf U}_4, {\bf U}_i^H{\bf U}_i = {\bf I}
\end{aligned}
\label{DLR_model.eqn}
\end{equation} 
where $\boldsymbol{{\cal S}}$ denotes the tail of Gaussian noise, $\mu_1$ and $\mu_2$ are two nonnegative parameters used to get a trade-off between the three terms of the objective function, and ${\Vert \boldsymbol{{\cal Z}} \Vert}_{\rm SCPD}$ denotes the structured CP tensor decomposition operation which will be expanded later. We can find that problem \eqref{DLR_model.eqn} is a nonconvex optimization problem due to the nonconvexity of tensor decomposition. Thus, we use the ADMM method \cite{bertsekas2014constrained} to solve problem \eqref{DLR_model.eqn} in this paper.

First of all, through introducing auxiliary variable $\boldsymbol{{\cal R}}$, the problem \eqref{DLR_model.eqn} can be rewritten as 

\noindent
\begin{equation} 
\begin{aligned}
&\min_{\boldsymbol{{\cal Z}},\boldsymbol{{\cal S}},\boldsymbol{{\cal R}}}{\Vert \boldsymbol{{\cal Y}} - \boldsymbol{{\cal Z}} -\boldsymbol{{\cal S}}\Vert}_F^2+\mu_1{\Vert \boldsymbol{{\cal Y}} - \boldsymbol{{\cal R}}-\boldsymbol{{\cal S}} \Vert}_F^2+\mu_2{\Vert \boldsymbol{{\cal S}} \Vert}_1 \\
&s.t. \quad \boldsymbol{{\cal Z}}=\boldsymbol{{\cal G}} \times_1 {\bf U}_1 \times_2 {\bf U}_2 \times_3 {\bf U}_3 \times_4 {\bf U}_4, {\bf U}_i^H{\bf U}_i = {\bf I}\\
&~~~~~~~ \boldsymbol{{\cal R}} = \sum_{l=1}^{L} {\bf a}_l \circ {\bf b}_l \circ {\bf c}_l \circ {\bf d}_l, \boldsymbol{{\cal Z}} = \boldsymbol{{\cal R}}
\end{aligned}
\label{DLR_model_rewritten.eqn}
\end{equation} 

Then, the augmented Lagrangian function of problem \eqref{DLR_model_rewritten.eqn} can be represented as 
\begin{equation} 
\begin{aligned}
&\min_{\boldsymbol{{\cal Z}},\boldsymbol{{\cal S}},\boldsymbol{{\cal R}}}{\Vert \boldsymbol{{\cal Y}} - \boldsymbol{{\cal Z}} -\boldsymbol{{\cal S}}\Vert}_F^2+\mu_1{\Vert \boldsymbol{{\cal Y}} - \boldsymbol{{\cal R}}-\boldsymbol{{\cal S}} \Vert}_F^2+\mu_2{\Vert \boldsymbol{{\cal S}} \Vert}_1 \\
&~~~~~~~~+ \langle\boldsymbol{\Lambda},\boldsymbol{{\cal Z}} - \boldsymbol{{\cal R}}\rangle + \frac{\gamma}{2} \Vert \boldsymbol{{\cal Z}} - \boldsymbol{{\cal R}} \Vert_F^2 \\
&s.t. \quad \boldsymbol{{\cal Z}}=\boldsymbol{{\cal G}} \times_1 {\bf U}_1 \times_2 {\bf U}_2 \times_3 {\bf U}_3 \times_4 {\bf U}_4, {\bf U}_i^H{\bf U}_i = {\bf I}\\
&~~~~~~~ \boldsymbol{{\cal R}} = \sum_{l=1}^{L} {\bf a}_l \circ {\bf b}_l \circ {\bf c}_l \circ {\bf d}_l
\end{aligned}
\label{DLR_model_Lagrange.eqn}
\end{equation} 
where $\gamma$ and $\boldsymbol{\Lambda} \in \mathbb{C}^{N_s N_b \times N_{st}\times K} \times M$ are the penalty parameter and Lagrange multiplier, respectively. Accordingly, problem \eqref{DLR_model_Lagrange.eqn} can be solved alternately by optimizing one variable while fixing others. Therefore, in the $(k+1)$-th iteration, variables in the problem \eqref{DLR_model_Lagrange.eqn} can be updated specifically as follows:

1) Update $\boldsymbol{{\cal Z}}$: Extracting all terms containing $\boldsymbol{{\cal Z}}$ from \eqref{DLR_model_Lagrange.eqn}, we can deduce that
\begin{equation} 
\begin{aligned}
&\boldsymbol{{\cal Z}}^{(k+1)}= \underset{\boldsymbol{{\cal Z}}}{\rm argmin} \; \ell (\boldsymbol{{\cal Z}},\boldsymbol{{\cal R}}^{(k)},\boldsymbol{{\cal S}}^{(k)},\boldsymbol{{\Lambda}}^{(k)}) \\
&= \underset{\substack{{{\boldsymbol{{\cal Z}}=\boldsymbol{{\cal G}} \times_1 {\bf U}_1 \times_2 {\bf U}_2 \times_3 {\bf U}_3 \times_4 {\bf U}_4}  } \\ {{\bf U}_i^H {\bf U}_i} = {\bf I}} } {\rm argmin} \; {\Vert \boldsymbol{{\cal Y}} - \boldsymbol{{\cal Z}} -\boldsymbol{{\cal S}}^{(k)} \Vert}_F^2 + \\
&~~~~\langle\boldsymbol{\Lambda}^{(k)},\boldsymbol{{\cal Z}} - \boldsymbol{{\cal R}}^{(k)}\rangle + \frac{\gamma}{2} \Vert \boldsymbol{{\cal Z}} - \boldsymbol{{\cal R}}^{(k)} \Vert_F^2
\end{aligned}
\label{update_P.eqn}
\end{equation}
which can be converted into the following equivalent problem:
\begin{equation} 
\begin{aligned}
&\underset{{\bf U}_i^H {\bf U}_i = {\bf I} } {\rm argmin} \; \Vert \boldsymbol{{\cal G}} \times_1 {\bf U}_1 \times_2 {\bf U}_2 \times_3 {\bf U}_3 \times_4 {\bf U}_4 - \frac{1}{2+\gamma}(2 \boldsymbol{{\cal Y}}  \\
& ~~~~~~~~~~~~-2 \boldsymbol{{\cal S}}^{(k)}+ \gamma \boldsymbol{{\cal R}}^{(k)} - \boldsymbol{\Lambda}^{(k)}) \Vert_F^2 
\end{aligned}
\label{update_P_eq.eqn}
\end{equation}
We resort to the classical HOOI algorithm \cite{kolda2009tensor} to solve problem \eqref{update_P_eq.eqn}, which can easily obtain $\boldsymbol{{\cal G}}^{(k+1)}$ and ${\bf U}_i^{(k+1)}$ with desired rank $[r_1,r_2,r_3,r_4]$. Accordingly, $\boldsymbol{{\cal Z}}$ can be updated by
\begin{equation} 
\begin{aligned}
 \boldsymbol{{\cal Z}}^{(k+1)} = \boldsymbol{{\cal G}}^{(k+1)} \times_1 {\bf U}_1^{(k+1)} \times_2 {\bf U}_2^{(k+1)} \times_3 {\bf U}_3^{(k+1)} \times_4 {\bf U}_4^{(k+1)} 
\end{aligned}
\label{update_P_sol.eqn}
\end{equation}

2) Update $\boldsymbol{{\cal R}}$: Extracting all terms containing $\boldsymbol{{\cal R}}$ from \eqref{DLR_model_Lagrange.eqn}, we can deduce that
\begin{equation} 
\begin{aligned}
&\boldsymbol{{\cal R}}^{(k+1)}= \underset{\boldsymbol{{\cal R}}}{\rm argmin} \; \ell (\boldsymbol{{\cal Z}}^{(k+1)},\boldsymbol{{\cal R}},\boldsymbol{{\cal S}}^{(k)},\boldsymbol{{\Lambda}}^{(k)}) \\
&= \underset{\boldsymbol{{\cal R}} = \sum_{l=1}^{L} {\bf a}_l \circ {\bf b}_l \circ {\bf c}_l \circ {\bf d}_l } {\rm argmin} \; \mu_1{\Vert \boldsymbol{{\cal Y}} - \boldsymbol{{\cal R}}-\boldsymbol{{\cal S}}^{(k)} \Vert}_F^2 + \langle\boldsymbol{\Lambda}^{(k)},\\
&~~~~\boldsymbol{{\cal Z}}^{(k+1)} - \boldsymbol{{\cal R}}\rangle + \frac{\gamma}{2} \Vert \boldsymbol{{\cal Z}}^{(k+1)} - \boldsymbol{{\cal R}} \Vert_F^2 \\
& = \underset{\boldsymbol{{\cal R}} = \sum_{l=1}^{L} {\bf a}_l \circ {\bf b}_l \circ {\bf c}_l \circ {\bf d}_l } {\rm argmin} \; \Vert \boldsymbol{{\cal R}} - \frac{1}{2\mu_1 + \gamma} (2 \mu_1 \boldsymbol{{\cal Y}} - 2 \mu_1 \boldsymbol{{\cal S}}^{(k)} \\
&~~~~+ \gamma \boldsymbol{{\cal Z}}^{(k+1)} + \boldsymbol{\Lambda}^{(k)}) \Vert_F^2 \\
&= \underset{\boldsymbol{{\cal R}}=\sum_{l=1}^{L} {\bf a}_l \circ {\bf b}_l \circ {\bf c}_l \circ {\bf d}_l}{\rm argmin} \; \Vert \boldsymbol{{\cal O}}^{(k+1)} - \boldsymbol{{\cal R}} \Vert_F^2
\end{aligned}
\label{update_Q.eqn}
\end{equation}
where $\boldsymbol{{\cal O}}^{(k+1)} = \frac{1}{2\mu_1 + \gamma} (2 \mu_1 \boldsymbol{{\cal Y}} - 2 \mu_1 \boldsymbol{{\cal S}}^{(k)}+ \gamma \boldsymbol{{\cal Z}}^{(k+1)} + \boldsymbol{\Lambda}^{(k)})$. It can be easily found that problem \eqref{update_Q.eqn} is a classical CPD problem. Thus, we can convert solving $\boldsymbol{{\cal R}}$ into solving factor matrices from $\boldsymbol{{\cal O}}^{(k+1)}$, i.e.
\begin{equation} 
\begin{aligned}
&{\bf A}^{(k+1)},{\bf B}^{(k+1)},{\bf C}^{(k+1)},{\bf D}^{(k+1)} = \\
&\underset{{\bf A},{\bf B},{\bf C},{\bf D} } {\rm argmin} \;  \left\Vert
\boldsymbol{{\cal O}}^{(k+1)} - \sum_{l=1}^{L} {\bf a}_l \circ {\bf b}_l \circ {\bf c}_l \circ {\bf d}_l   \right\Vert^2_F
\end{aligned}
\label{update_Q_eq.eqn}
\end{equation}

Although problem \eqref{update_Q_eq.eqn} can be solved by traditional ALS algorithm \cite{10552118}, this method is sensitive to noise, and the accuracy of the solution is seriously affected by the initialization \cite{9835123}. To overcome these shortcomings, motivated by the structured CPD method proposed in \cite{9835123}, we explore the 4D structured tensor decomposition channel estimation (4DSTDCE) method by exploiting the Vandermonde structure of the factor matrix, which achieves better robustness and accuracy than the ALS method without initialization and iteration.  Specifically, we first denote  $\boldsymbol{{\cal Q}} = {\rm Vec}^1_2(\boldsymbol{{\cal O}}) =  [\![{\bf B}\odot{\bf A},{\bf C}, {\bf D}]\!] \in \mathbb{C}^{N_s N_b  N_{st}\times K\times M}$ and ${\bf E} = {\bf B} \odot {\bf A} = [{\bf e}_1,\cdots,{\bf e}_L]$. Notice that we omit the representation of the $(k+1)$-th iteration in the following derivation. Then, problem \eqref{update_Q_eq.eqn} can be converted into the following equivalent problem:
\begin{equation}
\begin{aligned}
{\bf C},{\bf D},{\bf E} = \underset{{\bf C},{\bf D},{\bf E} } {\rm argmin} \;  \left\Vert
\boldsymbol{{\cal Q}} - \sum_{l=1}^{L} {\bf e}_l \circ {\bf c}_l \circ {\bf d}_l   \right\Vert^2_F
\end{aligned}
\label{three_order_CP_Decomposition_Problem.eqn}
\end{equation} 

Performing mode-$1$ unfolding on the tensor $\boldsymbol{{\cal Q}}$, we can obtain
\begin{equation}
\begin{aligned}
{\bf Q}^T_{(1)}= ({\bf D} \odot {\bf C}){\bf E}^T \in \mathbb{C}^{MK \times N_{st} N_s N_b}
\end{aligned}
\label{mode1_three_order_CPD.eqn}
\end{equation} 

Then, defining an integer pair $(K_4,L_4)$ subjected to $K_4+L_4=M+1$ and introducing the spatial smoothing transformation in \cite{9049103}, the spatial smoothing of ${\bf Q}_{(1)}^T$ can be given by
\begin{equation}
\begin{aligned}
{\bf Q}_s&= [({\bf M}_1 \otimes {\bf I}_K){\bf Q}_{(1)}^T, \cdots, ({\bf M}_{L_4} \otimes {\bf I}_K){\bf Q}_{(1)}^T] \\
&=[(({\bf M}_1 {\bf D}) \odot {\bf C}){\bf E}^T,\cdots,(({\bf M}_{L_4} {\bf D}) \odot {\bf C}){\bf E}^T] 
\end{aligned}
\label{R_s.eqn}
\end{equation} 
where ${\bf M}_l=[{\bf 0}_{K_4\times (l-1)},{\bf I}_{K_4},{\bf 0}_{K_4\times (L_4-l)}] \in \mathbb{C}^{K_4 \times M}$. 
Due to the Vandermonde structure of matrix $\bf D$, we can obtain ${\bf M}_l {\bf D} = [{\bf D}]_{(1:K_4,:)} {\rm diag}_{l-1}({\bf D})$, where $[{\bf D}]_{(1:K_4,:)}$ denotes the first $K_4$ rows of $\bf D$. Note that when $l=1$, ${\rm diag}_{l-1}({\bf D})$, i.e. ${\rm diag}_{0}({\bf D})$, is an identity matrix. Thus, \eqref{R_s.eqn} can be rewritten as
\begin{equation}
\begin{aligned}
{\bf Q}_s= &[([{\bf D}]_{(1:K_4,:)} {\rm diag}_{0}({\bf D})) \odot {\bf C}){\bf E}^T,\cdots,\\
&([{\bf D}]_{(1:K_4,:)} {\rm diag}_{L_4-1}({\bf D}))\odot {\bf C}){\bf E}^T] \\
=&([{\bf D}]_{(1:K_4,:)} \odot {\bf C}) ([{\bf 1}_{1 \times L};[{\bf D}]_{(1:L_4-1,:)}] \odot {\bf E})^T  
\end{aligned}
\label{R_s_smooth_ultimate.eqn}
\end{equation} 
where ${\bf 1}_{1\times L}$ denotes a row vector with $L$ entries that are all $1$. Then, we perform the truncated singular value decomposition (tSVD) on matrix ${\bf Q}_s$ with rank $L$, yielding ${\bf Q}_s = {\bf U}_s {\bf \Sigma} {\bf V}_s^H$. Since $[{\bf D}]_{(1:K_4,:)} \odot {\bf C}$ is full column rank matrix and assuming the number of paths $L$ is known, there exists a full rank matrix ${\bf J} \in \mathbb{C}^{L\times L}$, satisfying
\begin{equation}
\begin{aligned}
&{\bf U}_s {\bf J} = [{\bf D}]_{(1:K_4,:)} \odot {\bf C} \\
&{\bf U}_s {\bf J}_{(:,l)}  =  [{\bf D}]_{(1:K_4,l)} \otimes {\bf c}_l\\
\end{aligned}
\label{R_s_SVD.eqn}
\end{equation}
\begin{equation}
\begin{aligned}
{\bf V}_s^* {\bf \Sigma} {\bf J}^{-T} =[{\bf 1}_{1 \times L};[{\bf D}]_{(1:L_4-1,:)}] \odot {\bf E}
\end{aligned}
\label{Right_singular_SVD.eqn}
\end{equation} 

Consequently, we have
\begin{equation}
\begin{aligned}
&{\bf U}^s_1 {\bf J} = [{\bf D}]_{(2:K_4,:)} \odot {\bf C} \\
&{\bf U}^s_2 {\bf J} = [{\bf D}]_{(1:K_4-1,:)} \odot {\bf C} = {\bf U}^s_1 {\bf J} {\boldsymbol{\Omega}}
\end{aligned}
\label{R_s_SVD_U1_U2.eqn}
\end{equation}
where ${\bf U}^s_1 = [{\bf U}_s]_{(1:K(K_4-1),:)}$, ${\bf U}^s_2 = [{\bf U}_s]_{(K+1:KK_4,:)}$, and ${\boldsymbol{\Omega}} = {\rm diag}(e^{j \omega_1},\cdots,e^{j \omega_L})$. It can be easily seen that the elements of ${\boldsymbol{\Omega}}$ are related to the first row of factor matrix $\bf D$. According to \eqref{R_s_SVD_U1_U2.eqn}, we can obtain 
\begin{equation}
\begin{aligned}
{{\bf U}^s_1}^{\dagger} {\bf U}^s_2 = {\bf J} \boldsymbol{\Omega} {\bf J}^{-1} = {\bf P \Lambda P}^{-1}
\end{aligned}
\label{EVD.eqn}
\end{equation}
where $\bf \Lambda$ and $\bf P$ are the eigenvalue and eigenvector matrices of ${{\bf U}^s_1}^{\dagger} {\bf U}^s_2$, respectively. Obviously, \eqref{EVD.eqn} is a similarity transformation, satisfying
\begin{equation}
\begin{aligned}
{\bf P} &= {\bf J } \boldsymbol{\Delta} {\bf \Pi} \\
{\bf \Lambda} &= {\boldsymbol{\Omega \Pi} } 
\end{aligned}
\label{similarity_transformation.eqn}
\end{equation}
where $\boldsymbol{\Delta}$ and $\bf \Pi$ are the unknown nonsingular diagonal scaling ambiguity matrix and permutation matrix, respectively. Thus, the permutation factor matrix $\widetilde{\bf D}$ can be estimated by
\begin{equation}
\begin{aligned}
\widetilde{\bf D} = {\bf D} {\bf \Pi} = [\tilde{{\bf d}}_1,\cdots,\tilde{{\bf d}}_L]
\end{aligned}
\label{D_obtained.eqn}
\end{equation}
where $\tilde{{\bf d}}_l = [{\bf \Lambda}_{(l,l)},\cdots, {\bf \Lambda}_{(l,l)}^{M}]^T$.

Then, according to \eqref{R_s_SVD.eqn} and \eqref{similarity_transformation.eqn}, we have
\begin{equation}
\begin{aligned}
&[{\bf U}_s]_{(:,1:L)} {\bf P} = {\bf U}_s{\bf J } \boldsymbol{\Delta} {\bf \Pi} = ([{\bf D}]_{(1:K_4,:)}  \odot {\bf C} ) \boldsymbol{\Delta} {\bf \Pi}\\
& =([{\bf D}]_{(1:K_4,:)} \odot ({\bf C} \boldsymbol{\Delta}))  {\bf \Pi} = ([{\bf D}]_{(1:K_4,:)}{\bf \Pi} )\odot ({\bf C} \boldsymbol{\Delta}  {\bf \Pi}) \\
&=[\widetilde{{\bf D}}]_{(1:K_4,:)} \odot \widetilde{{\bf C}} = [\widetilde{{\bf D}}]_{(1:K_4,:)} \odot [\tilde{{\bf c}}_1,\cdots,\tilde{{\bf c}}_L]
\end{aligned}
\label{UP_estimate.eqn}
\end{equation}
where $\widetilde{{\bf C}}$ is the permutation and scaling ambiguity matrix of $\bf C$. Thus, we can estimate the column vector $\tilde{{\bf c}}_l$ of the factor matrix $\widetilde{\bf C}$ by solving the following problem
\begin{equation}
\begin{aligned}
\tilde{{\bf c}}_l = \underset{{\bf c} _l} {\rm argmin} \;  \left\Vert  [\widetilde{{\bf D}}]_{(1:K_4,l)} \otimes {\bf c}_l -{\bf U}_s [{\bf P}]_{(:,l)} \right\Vert^2_F
\end{aligned}
\label{C_est.eqn}
\end{equation}
Apparently, problem \eqref{C_est.eqn} is a typical least squares problem, which admits a closed-form solution
\begin{equation}
\begin{aligned}
\tilde{{\bf c}_l} = \frac{[\widetilde{{\bf D}}]_{(1:K_4,l)}^H \otimes {\bf I}_K}{[\widetilde{{\bf D}}]_{(1:K_4,l)}^H [\widetilde{{\bf D}}]_{(1:K_4,l)}}  {\bf U}_s [{\bf P}]_{(:,l)}
\end{aligned}
\label{C_est_solution.eqn}
\end{equation}

Subsequently, by substituting \eqref{similarity_transformation.eqn} into \eqref{Right_singular_SVD.eqn}, we can derive
\begin{equation}
\begin{aligned}
&{\bf V}_s^* {\bf \Sigma} {\bf P}^{-T} = {\bf V}_s^* {\bf \Sigma} {\bf J}^{-T} \boldsymbol{\Delta}^{-T} {\bf \Pi}^{-T} \\
&=([{\bf 1}_{1 \times L};[{\bf D}]_{(1:L_4-1,:)}] \odot {\bf E})  \boldsymbol{\Delta}^{-T} {\bf \Pi} \\
&= ([{\bf 1}_{1 \times L};[{\bf D}]_{(1:L_4-1,:)}] {\bf \Pi})\odot ({\bf E} \boldsymbol{\Delta}^{-T} {\bf \Pi}) \\
&= [{\bf 1}_{1 \times L};[\widetilde{{\bf D}}]_{(1:L_4-1,:)}] \odot \widetilde{{\bf E}} = \widehat{{\bf D}} \odot [\tilde{{\bf e}}_1,\cdots,\tilde{{\bf e}}_L]
\end{aligned}
\label{Right_singular_using_Q.eqn}
\end{equation} 
where $\widetilde{{\bf E}}$ is the permutation and scaling ambiguity matrix of $\bf E$, and $\widehat{{\bf D}} =  [{\bf 1}_{1 \times L};[\widetilde{{\bf D}}]_{(1:L_4-1,:)}] $. Similar to \eqref{C_est.eqn} and \eqref{C_est_solution.eqn}, the column vector $\tilde{{\bf e}}_l$ of the factor matrix $\widetilde{\bf E}$ is given by
\begin{equation}
\begin{aligned}
\tilde{{\bf e}}_l = \frac{[\widehat{{\bf D}}]_{(:,l)}^H \otimes {\bf I}_{N_s N_b N_{st}}}{[\widehat{{\bf D}}]_{(:,l)}^H [\widehat{{\bf D}}]_{(:,l)}}   {\bf V}_s^* {\bf \Sigma} [\tilde{\bf P}]_{(:,l)}
\end{aligned}
\label{E_est_solution.eqn}
\end{equation}
where $\tilde{\bf P} = {\bf P}^{-T}$. Since $\tilde{{\bf e}}_l$ is obtained by the Kronecker product of two rank-one vectors, the permutation and scaling ambiguity matrices $\widetilde{\bf A} = [\tilde{{\bf a}}_1,\cdots,\tilde{{\bf a}}_L]$ and $\widetilde{\bf B} = [\tilde{{\bf b}}_1,\cdots,\tilde{{\bf b}}_L]$ can be estimated by solving a least-squares Khatri-Rao factorization problem, i.e. 
\begin{equation}
\begin{aligned}
\tilde{{\bf a}}_l, \tilde{{\bf b}}_l= \underset{{\bf a}_l,{\bf b}_l} {\rm argmin} \;  \left\Vert  \widetilde{{\bf E}}_l - {\bf a}_l {\bf b}_l^T \right\Vert^2_F
\end{aligned}
\label{AB_est_solution.eqn}
\end{equation}
where $\widetilde{{\bf E}}_l = {\rm unvec}_{N_s N_b \times N_{st}}(\tilde{{\bf e}}_l)$. Since $\tilde{{\bf a}}_l$ and $\tilde{{\bf b}}_l$ are rank-one vectors, problem \eqref{AB_est_solution.eqn} can be solved by performing SVD on $\widetilde{{\bf E}}_l$, obtaining the sorted diagonal singular value matrix $\hat{{\bf \Sigma}}^l$ and the corresponding left singular matrix $\hat{\bf U}^l$ and right singular matrix $\hat{\bf V}^l$ \cite{5530416}. Hence, $\tilde{{\bf a}}_l$ and $\tilde{{\bf b}}_l$ are estimated as $\tilde{{\bf a}}_l= [\hat{{\bf \Sigma}}^l]_{(1,1)} [\hat{\bf U}^l]_{(:,1)}$ and $\tilde{{\bf b}}_l=[\hat{\bf V}^l]_{(:,1)}^*$, respectively. 

Overall, the 4DSTDCE method is summarized in Algorithm \ref{4DSTDCE_method}. It is worth mentioning that owing to the small $L$ in mmWave systems, CPD can achieve unique decomposition by choosing appropriate system parameters. Due to space limitations, the proof for the uniqueness of quadrilinear tensor decomposition can use the theorem in \cite{quadrilinear_decomposition} for reference.
\begin{algorithm}[htbp]
	\caption{4DSTDCE Method}
	\label{4DSTDCE_method}
	\begin{algorithmic}  
		\State {\bf Inputs}: 
		4D tensor $\boldsymbol{{\cal O}}$.	
		\begin{itemize}

			\item[{\bf 1)}]  Define $(K_4,L_4)$ and calculate the spatial smoothing ${\bf Q}_s$ by \eqref{R_s.eqn}.
			
			\item[{\bf 2)}] Compute the SVD of ${\bf Q}_s$ as ${\bf Q}_s = {\bf U}_s {\bf \Sigma} {\bf V}_s^H$ and obtain ${\bf U}^s_1$ and ${\bf U}^s_2$.
			
			\item[{\bf 3)}] Estimate $\widetilde{\bf D}$ by \eqref{D_obtained.eqn}.
			
			\item[{\bf 4)}] Estimate $\widetilde{\bf C}$ by \eqref{C_est_solution.eqn}.
			
			\item[{\bf 5)}] Estimate $\widetilde{\bf B}$ and $\widetilde{\bf A}$ by \eqref{AB_est_solution.eqn}.
		\end{itemize}
		\State{\bf Output}: Estimated factor matrices  $\widetilde{\bf A}$, $\widetilde{\bf B}$, $\widetilde{\bf C}$, and $\widetilde{\bf D}$. 
	\end{algorithmic}
\end{algorithm}	

The permutation matrix $\bf \Pi$ is shared among all four factor matrices, which can be ignored when calculating $\boldsymbol{{\cal R}}^{(k+1)}$. Thus, after obtaining $\widetilde{\bf A}^{(k+1)}$, $\widetilde{\bf B}^{(k+1)}$, $\widetilde{\bf C}^{(k+1)}$, and $\widetilde{\bf D}^{(k+1)}$ from $\boldsymbol{{\cal R}}^{(k+1)}$, the solution to problem \eqref{update_Q.eqn} is given by
\begin{equation}
\begin{aligned}
\boldsymbol{{\cal R}}^{(k+1)}=\sum_{l=1}^{L} \tilde{{\bf a}}_l^{(k+1)} \circ \tilde{{\bf b}}_l^{(k+1)} \circ \tilde{{\bf c}}_l^{(k+1)} \circ \tilde{{\bf d}}_l^{(k+1)}
\end{aligned}
\label{solution_to_Q.eqn}
\end{equation}

3) Update $\boldsymbol{{\cal S}}$: Extracting all terms containing $\boldsymbol{{\cal S}}$ from \eqref{DLR_model_Lagrange.eqn}, we can deduce that
\begin{equation} 
\begin{aligned}
\boldsymbol{{\cal S}}^{(k+1)}&= \underset{\boldsymbol{{\cal S}}}{\rm argmin} \; \ell (\boldsymbol{{\cal Z}}^{(k+1)},\boldsymbol{{\cal R}}^{(k+1)},\boldsymbol{{\cal S}},\boldsymbol{{\Lambda}}^{(k)}) \\
&= \underset{\boldsymbol{{\cal S}}}{\rm argmin} \; {\Vert \boldsymbol{{\cal Y}} - \boldsymbol{{\cal Z}}^{(k+1)} -\boldsymbol{{\cal S}}\Vert}_F^2+\mu_1 \Vert \boldsymbol{{\cal Y}} - \\
&~~~~~~~~~~~~~	\boldsymbol{{\cal R}}^{(k+1)}-\boldsymbol{{\cal S}} \Vert_F^2+\mu_2{\Vert \boldsymbol{{\cal S}} \Vert}_1  \\
&=\underset{\boldsymbol{{\cal S}}}{\rm argmin} \; \frac{\mu_2}{1+\mu_1}  {\Vert \boldsymbol{{\cal S}} \Vert}_1 + \Vert \boldsymbol{{\cal S}} - (\boldsymbol{{\cal Y}} -\frac{1}{1+\mu_1}\\
&~~~~~~~~~~~~(\boldsymbol{{\cal Z}}^{(k+1)} + \mu_1 \boldsymbol{{\cal R}}^{(k+1)} ) ) \Vert_F^2
\end{aligned}
\label{update_S.eqn}
\end{equation}

Problem \eqref{update_S.eqn} can be solved by using the well-known soft-thresholding operator, which is defined as ${\cal R}_{\bigtriangledown}(\varpi) = {\rm sign}(\varpi)*{\rm max}(|\varpi| - \bigtriangledown/2,0)$. Accordingly, the solution to problem \eqref{update_S.eqn} can be expressed as
\begin{equation} 
\begin{aligned}
\boldsymbol{{\cal S}}^{(k+1)}= {\cal R}_{\frac{\mu_2}{1+\mu_1}}(\boldsymbol{{\cal Y}} - \frac{1}{1+\mu_1}(\boldsymbol{{\cal Z}}^{(k+1)} + \mu_1 \boldsymbol{{\cal R}}^{(k+1)}))
\end{aligned}
\label{solution_of_S.eqn}
\end{equation} 

4) Update multiplier $\boldsymbol{\Lambda}$: Giving $\boldsymbol{{\cal Z}}^{(k+1)}$ and $\boldsymbol{{\cal R}}^{(k+1)}$, the multiplier $\boldsymbol{\Lambda}$ is updated by the following equations:
\begin{equation} 
\begin{aligned}
\boldsymbol{{\Lambda}}^{(k+1)}=\boldsymbol{{\Lambda}}^{(k)}+\gamma(\boldsymbol{{\cal Z}}^{(k+1)} -  \boldsymbol{{\cal R}}^{(k+1)} )
\end{aligned}
\label{update_multipliers}
\end{equation} 	

Overall, the proposed DLR4DTD method is summarized in Algorithm \ref{DLR4DTD_method}.
\begin{algorithm}[htbp]
	\caption{DLR4DTD Method}
	\label{DLR4DTD_method}
	\begin{algorithmic}  
		\State {\bf Inputs}: 
		Observed signal $\boldsymbol{{\cal Y}}$, parameters $\mu_1= 0.01$ and $\mu_2=0.5$, and the number of paths $L$.	
		
		\State {\bf Initialization}:
		Optimization terms $\boldsymbol{{\cal Z}}=\boldsymbol{{\cal R}} = \boldsymbol{{\cal S}} = \boldsymbol{0}$, multiplier $\boldsymbol{{\Lambda}}=0$, penalty parameter $\gamma = 1.5$, iteration $k = 1$, maximum iteration $k_{max} = 300$, tolerance $\varepsilon_1 = \varepsilon_2 = \varepsilon_3 = 10^{-5}$, ${\rm Obj} = {\Vert \boldsymbol{{\cal Y}} - \boldsymbol{{\cal Z}} -\boldsymbol{{\cal S}}\Vert}_F^2+\mu_1{\Vert \boldsymbol{{\cal Y}} - \boldsymbol{{\cal R}}-\boldsymbol{{\cal S}} \Vert}_F^2+\mu_2{\Vert \boldsymbol{{\cal S}} \Vert}_1$, and $r_1 = r_2 = r_3 = r_4 = L$. 
		
		\State{\bf while} not meet the stopping condition {\bf do}
		\begin{itemize}
			\item[] 
		\begin{itemize}	
			\item[{\bf 1)}] Update $\boldsymbol{{\cal Z}}$ by \eqref{update_P_sol.eqn}.
			
			\item[{\bf 2)}] Update $\boldsymbol{{\cal R}}$ by Algorithm \ref{4DSTDCE_method} and \eqref{solution_to_Q.eqn}, and obtain factor matrices  $\widetilde{\bf A}$, $\widetilde{\bf B}$, $\widetilde{\bf C}$, and $\widetilde{\bf D}$.
			
			\item[{\bf 3)}] Update $\boldsymbol{{\cal S}}$ by \eqref{solution_of_S.eqn}.
			
			\item[{\bf 4)}] Update $\boldsymbol{\Lambda}$ by \eqref{update_multipliers}.
			
			\item[{\bf 5)}] Update ${\rm Obj}$.
			
			\item[{\bf 6)}] Check the iteration stopping condition:
			\begin{itemize}
				\item[] $(k<k_{max}) ~~||~~ ({\rm max}({\rm vec}(\boldsymbol{{\cal Z}}^{(k+1)} - \boldsymbol{{\cal Z}}^{(k)})) < \varepsilon_1 ~~\&\&~~  {\rm max}({\rm vec}(\boldsymbol{{\cal R}}^{(k+1)} - \boldsymbol{{\cal R}}^{(k)})) < \varepsilon_2)  ~~||~~ ({\rm Obj}^{(k)}-{\rm Obj}^{(k+1)})<\varepsilon_3$
			\end{itemize}
			\item[{\bf 7)}] Update $k=k+1$.
		\end{itemize}
	\end{itemize}
		\State{\bf end while} 
		\State{\bf Output}: Estimated factor matrices  $\widetilde{\bf A}$, $\widetilde{\bf B}$, $\widetilde{\bf C}$, and $\widetilde{\bf D}$. 
	\end{algorithmic}
\end{algorithm}	

\subsection{Channel Parameters Estimation}
Based on the estimated factor matrices, we can extract the channel parameters. As mentioned before, the permutation matrix $\bf \Pi$ is shared among all four factor matrices. Thus, the channel parameters of each path are automatically paired, which does not need to consider the parameters matching problem. The estimation of channel parameters corresponding to each path is detailed as follows:
\subsubsection{Estimate $\phi_{\rm BR}$ and $\theta_{\rm RM}^l$}
Parameters $\phi_{\rm BR}$ and $\theta_{\rm RM}^l$ are embedded in factor matrix $\bf A$. Although the $L$ paths share the same ${\tilde{\phi}}_{\rm BR}^l$, we will obtain different estimated ${\tilde{\phi}}_{\rm BR}^l$ values in different paths. By designing correlation-based estimator, parameters ${\tilde{\phi}}_{\rm BR}^l$ and $\tilde{\theta}_{\rm RM}^l$ can be estimated by
\begin{equation}
\begin{aligned}
{\tilde{\phi}}_{\rm BR}^l,\tilde{\theta}_{\rm RM}^l =\underset{\phi_{\rm BR}^l,\theta_{\rm RM}^l}{\rm argmax} \; \frac{\vert \tilde{{\bf a}}_l^H \boldsymbol{\Upsilon}^T {\bf a}_{s}(\phi_{\rm BR}^l,\theta_{\rm RM}^l) \vert}{\Vert \tilde{{\bf a}}_l \Vert_2 \Vert\boldsymbol{\Upsilon}^T {\bf a}_{s}(\phi_{\rm BR}^l,\theta_{\rm RM}^l)\Vert_2}
\end{aligned}
\label{AoA.eqn}
\end{equation}

Problem \eqref{AoA.eqn} can be solved through a two-dimensional search method. However, the estimation accuracy of the search-based method is limited by the precision of grid partitioning. The more precise the two-dimensional grid division, the greater the computational complexity. Thus, we propose a two-stage two parameters estimation method to reduce computational complexity while ensuring estimation accuracy. Specifically, in the first stage, a rough two-dimension search with larger grid spacing is used to obtain the initial solution of $\hat{\phi}_{\rm BR}^l$ and $\hat{\theta}_{\rm RM}^l$. In the second stage, a Nelder-Mead simplex method \cite{nelder1965simplex} is employed to obtain the exact solution $\tilde{\theta}_{\rm RM}^l$ and $\tilde{\phi}_{\rm BR}^l$ under the initial solution obtained in the first stage. The Nelder-Mead simplex method is an iterative nonlinear optimization method that can achieve rapid convergence without any derivative information \cite{lagarias1998convergence}. Notice that a suitable initial solution is the key to the method. If the initial solution is chosen improperly, it may cause the algorithm to stagnate near the local optimum or converge slowly. Thus, the proposed method combines the two-stage and can get a good trade-off between complexity and accuracy. 

Finally, the final estimated ${\tilde{\phi}}_{\rm BR}$ is calculated by averaging the ${\tilde{\phi}}_{\rm BR}^l$ obtained from different paths.

\subsubsection{Estimate $\theta_{\rm BR}$ and $\phi_{\rm RM}^l$}
Parameters $\theta_{\rm BR}$ and $\phi_{\rm RM}^l$ are embedded in factor matrix $\bf B$. Similar to $\phi_{\rm BR}$, we will obtain different estimated $\theta_{\rm BR}^l$ values in different paths. For each path, we have $\tilde{\bf b}_l = \varkappa_l \boldsymbol{\Xi}^T \boldsymbol{\Theta} {\bf J}_d(\tilde{\theta}_{eq}^l) {\bf v}(\tilde{\phi}_{eq}^l) + {\bf n}_l $, where $\varkappa_l$ and ${\bf n}_l$ denote the unknown scaling ambiguity scalar of path $l$ and the error vector, respectively. By leveraging the orthogonality between signal subspace and noise subspace and denoting $\boldsymbol{\Xi}_{\Theta} = \boldsymbol{\Xi}^T \boldsymbol{\Theta}$, we can obtain the spectrum function
\begin{equation}
\begin{aligned}
&f({{\theta}}_{\rm BR}^l,{\phi}_{\rm RM}^l) = \\
&\frac{1}{{\bf v}({\phi}_{eq}^l)^H {\bf J}_d({\theta}_{eq}^l) \boldsymbol{\Xi}_{\Theta}^H {\bf E}_n^l ({\bf E}_n^l )^H \boldsymbol{\Xi}_{\Theta}  {\bf J}_d({\theta}_{eq}^l) {\bf v}({\phi}_{eq}^l)}
\end{aligned}
\label{Music.eqn}
\end{equation}
where ${\bf E}_n^l$ is calculated by the eigenvector matrix formed by the eigenvectors corresponding to the eigenvalues other than the maximum eigenvalue of $\tilde{b}_l \tilde{b}_l^H$. Then, the spectral peak search is employed to find the spectral peak and determine the parameters ${\hat{\theta}}_{\rm BR}^l$ and $\hat{\phi}_{\rm RM}^l$. Due to the fact that spectral peak search is also based on grid search, it is also affected by the precision of grid division. Considering this, we use the Nelder-Mead simplex method to improve the estimation performance by solving the following problem:
\begin{equation}
\begin{aligned}
{\tilde{\theta}}_{\rm BR}^l,\tilde{\phi}_{\rm RM}^l =\underset{\phi_{\rm BR}^l,\theta_{\rm RM}^l}{\rm argmax} \; \frac{\vert \tilde{{\bf b}}_l^H \boldsymbol{\Xi}_{\Theta}  {\bf J}_d(\tilde{\theta}_{eq}^l) {\bf v}(\tilde{\phi}_{eq}^l) \vert}{\Vert \tilde{{\bf b}}_l \Vert_2 \Vert\boldsymbol{\Xi}_{\Theta}  {\bf J}_d(\tilde{\theta}_{eq}^l) {\bf v}(\tilde{\phi}_{eq}^l)\Vert_2}
\end{aligned}
\label{AoD.eqn}
\end{equation}
where the initial solution is calculated by the spectral peak search solution to problem \eqref{Music.eqn}.

Finally, similar to $\phi_{\rm BR}$, the ${\tilde{\theta}}_{\rm BR}^l$ obtained from different paths are averaged as the final estimated ${\tilde{\theta}}_{\rm BR}$.
\subsubsection{Estimate $\tau_l$}
Parameter $\tau_l$ is embedded in factor matrix $\bf C$, which can be estimated by solving the following problem:
\begin{equation}
\begin{aligned}
\tilde{\tau}_l = \underset{\tau_l}{\rm argmax} \; \frac{\vert \tilde{{\bf c}}_l^H {\bf g}(\tau_l) \vert}{\Vert \tilde{{\bf c}}_l \Vert_2 \Vert {\bf g}(\tau_l) \Vert_2}
\end{aligned}
\label{time_delay_solution.eqn}
\end{equation}
Problem \eqref{time_delay_solution.eqn} can be solved by a one-dimensional search with high grid division precision. It is worth mentioning that the estimated $\tilde{\tau}_l$ is cascade time delay, which can be combined with $\phi_{\rm BR}$, $\theta_{\rm RM}^l$,  $\theta_{\rm BR}$, and $\phi_{\rm RM}^l$ to solve for the position of RIS, equivalent scatterers, and MS. Thus, if we want to decouple the cascade time delay, we can establish $3L$ equations containing $2L+2$ unknowns to estimate these positions firstly. The equations can be solved when $L\geq 2$. Then, we can calculate the distance of each path based on the estimated positions. Accordingly, the cascade time delay can be decoupled by leveraging the relationship between time, distance, and velocity.
\subsubsection{Estimate $f_d^l$}
Parameter $f_d^l$ is embedded in factor matrix $\bf D$, which can be directly estimated by \eqref{D_obtained.eqn}, detailed as
\begin{equation}
\begin{aligned}
\tilde{f}_d^l = \frac{1}{2\pi T_s N_b N_{st}} \angle({\bf \Lambda}_{(l,l)})
\end{aligned}
\label{fl_solution.eqn}
\end{equation}
where $\angle(\cdot)$ denotes the phase extraction operation.

\subsubsection{Estimate $\rho_l$}
After obtaining the channel parameters estimated before, we can calculate the permutation ambiguity cascade channel gain at the $k$-th pilot subcarrier in the $m$-th aggregated slot $\boldsymbol{\tilde{\rho}}_{k,m} =[\tilde{\rho}_{1,k,m},\cdots,\tilde{\rho}_{L,k,m}]^T$  by solving the following least-squares problem:
\begin{equation}
\begin{aligned}
\tilde{\rho}_{l,k,m} =\underset{\rho_l}{\rm argmin} \; &\Vert {\bf Y}_{k,m} -  \sum_{l=1}^{L} \rho_l \boldsymbol{\Phi}_{l,k,m} \Vert_F^2 
\end{aligned}
\label{LS_rewritten.eqn}
\end{equation}
where $\boldsymbol{\Phi}_{l,k,m} = e^{-j2 \pi \frac{k}{N} f_s \tilde{\tau}_l}\boldsymbol{\Upsilon}^T {\bf a}_{s}(\tilde{\phi}_{\rm BR},\tilde{\theta}_{\rm RM}^l) {\bf a}_{r}^T(\tilde{\theta}_{\rm BR},\tilde{\phi}_{\rm RM}^l)$ $\boldsymbol{\Xi}e^{j \tilde{\omega}_l m}$, $\tilde{\omega} = 2 \pi \tilde{f}_d^l T_s N_b N_{st}$. Defining matrix $\boldsymbol{\Gamma}_{k,m}$ whose entries $\boldsymbol{\Gamma}_{k,m}^{(i,j)}$ are equal to ${\rm Tr}(\boldsymbol{\Phi}_{j,k,m} \boldsymbol{\Phi}_{i,k,m}^H)$, the solution to problem \eqref{LS_rewritten.eqn} can be expressed as
\begin{equation}
\begin{aligned}
\boldsymbol{\tilde{\rho}}_{k,m} = \boldsymbol{\Gamma}_{k,m}^{-1} \boldsymbol{\zeta}_{k,m}
\end{aligned}
\label{Gamma_solution.eqn}
\end{equation}
where $\boldsymbol{\zeta}_{k,m} = $ $[{\rm Tr}({\bf Y}_{k,m} \boldsymbol{\Phi}_{1,k,m}^H),\cdots,{\rm Tr}({\bf Y}_{k,m} \boldsymbol{\Phi}_{L,k,m}^H)]^T$. Then, by averaging all the solutions $\boldsymbol{\tilde{\rho}}_{k,m}$ for $k = 1,\cdots, K$ and $m = 1,\cdots,M$, we can obtain the permutation ambiguity cascade channel gain $\boldsymbol{\tilde{\rho}} = [\tilde{\rho}_1,\cdots,\tilde{\rho}_L]^T$.

\section{CRB Analysis} \label{CRB_ana}	
CRB provides a lower bound on the variance of any unbiased estimator for a parameter, indicating the theoretical limit of estimation accuracy \cite{kay1993fundamentals}. Thus, in this section, we derive the CRB for estimated channel parameters $\{\tilde{\phi}_{\rm BR},\tilde{\theta}_{\rm RM}^l, \tilde{\theta}_{\rm BR}, \tilde{\phi}_{\rm RM}^l, \tilde{\tau_l}, \tilde{f}_d^l, \tilde{\rho}_l\}$ as a benchmark to evaluate the performance of the proposed method. To overcome the complexity of the CRB derivation process based on the mode-$n$ unfolding of the received signal, such as the derivation in \cite{7914672}, we propose a more concise derivation of the CRB by vectorizing the received signal. Specifically, for the convenience of deduction, we first let $\boldsymbol{\theta}_{\rm RM} = [{\theta}_{\rm RM}^1,\cdots,{\theta}_{\rm RM}^L]^T$, $\boldsymbol{\phi}_{\rm RM} = [{\phi}_{\rm RM}^1, \cdots,{\phi}_{\rm RM}^L]^T$, $\boldsymbol{\tau} = [\tau_1,\cdots,\tau_L]^T$, ${\bf f}_d=[f_d^1,\cdots,f_d^L]^T$, $\boldsymbol{\rho}=[\rho_1,\cdots,\rho_L]^T$, and $\boldsymbol{\eta} = [{\phi}_{\rm BR}, \boldsymbol{\theta}_{\rm RM}^T,{\theta}_{\rm BR}, \boldsymbol{\phi}_{\rm RM}^T,\boldsymbol{\tau}^T,{\bf f}_d^T,\boldsymbol{\rho}^T]^T$. Then, by vectoring the mode-$1$ unfolding of the noisy received signal $\boldsymbol{ {\cal Y}}$ and received signal $\boldsymbol{{\cal Z}}$ of \eqref{tensor_recieved_signal_modenpruduct.eqn}, we have ${\bf y}_{(1)} = {\rm vec}({\bf Y}_{(1)})$ and ${\bf z}_{(1)} = {\rm vec}({\bf Z}_{(1)})$. Accordingly, the log-likelihood function of $\boldsymbol{\eta}$ is derived as 
\begin{equation}
\begin{aligned}
{\cal L}(\boldsymbol{\eta}) = c - ({\bf y}_{(1)} - {\bf z}_{(1)})^H {\bf C}_{\boldsymbol{ {\cal N}} }^{-1} ({\bf y}_{(1)} - {\bf z}_{(1)})
\end{aligned}
\label{log-likelihood_function.eqn}
\end{equation}
where $c$ is a constant unrelated to $\boldsymbol{\eta}$, and ${\bf C}_{\boldsymbol{ {\cal N}}}$ is the covariance matrix of the noise vector ${\rm vec}({\bf N}^{\bf W}_{(1)})$. 

Furthermore, the complex Fisher information matrix (FIM) can be calculated by 
\begin{equation}
\begin{aligned}
{\rm FIM}(\boldsymbol{\eta}) = {\mathbb E} \left\{ \left(\frac{\partial {\cal L}(\boldsymbol{\eta})}{\partial \boldsymbol{\eta}}\right)^H  \frac{\partial {\cal L}(\boldsymbol{\eta})}{\partial \boldsymbol{\eta}} \right\}
\end{aligned}
\label{partial_derivation_log-likelihood_function.eqn}
\end{equation}

Finally, the CRB for $\boldsymbol{\eta}$ can be obtained by 
\begin{equation}
\begin{aligned}
{\rm CRB}(\boldsymbol{\eta}) = {\rm FIM}^{-1}(\boldsymbol{\eta}) 
\end{aligned}
\label{CRB.eqn}
\end{equation}
 
The concise closed-form expression for FIM is derived in detail in Appendix \ref{CRB_deviation}. 

\section{Simulation Results} \label{simulation_results}
\begin{figure*}[b]
	\centering
	\subfigure{
		\begin{minipage}[h]{0.24\linewidth}
			\centering
			\includegraphics[width=1.7in]{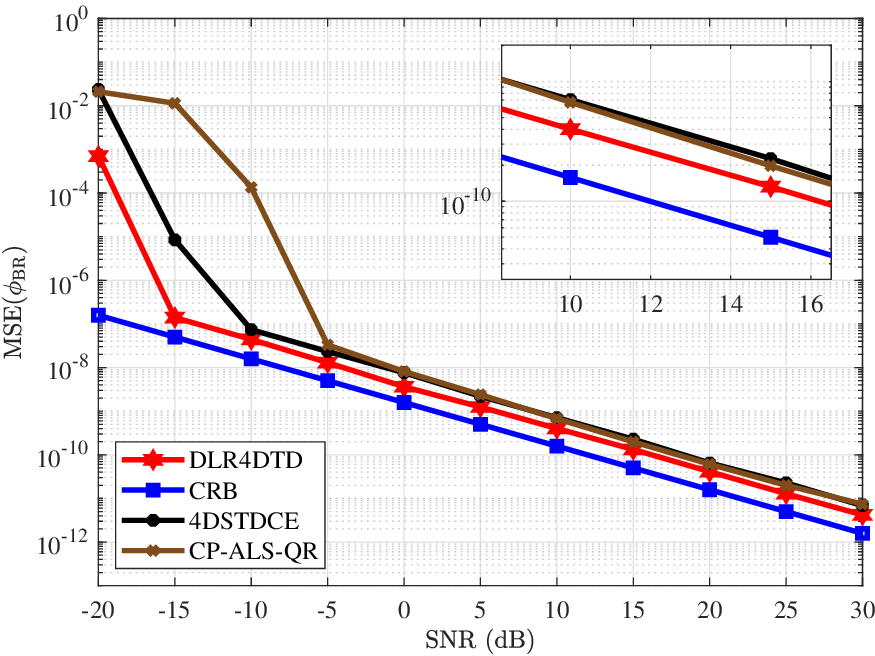}		
			\caption*{(a) $\rm MSE$ of ${\phi}_{\rm BR}$}
			\label{F1_phi_BR}
		\end{minipage}%
	}%
	\subfigure{
		\begin{minipage}[h]{0.24\linewidth}
			\centering
			\includegraphics[width=1.7in]{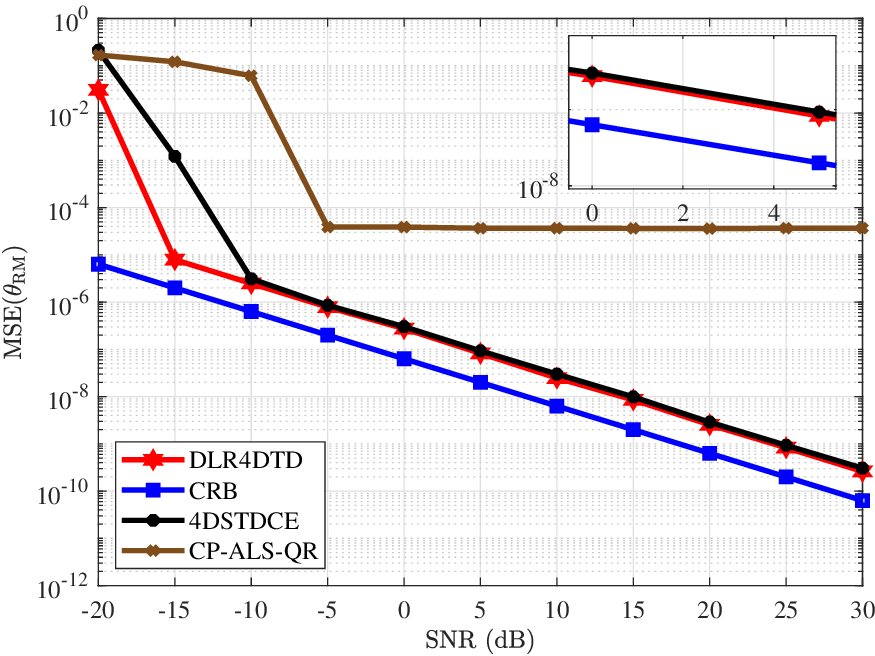}
			\caption*{(b) $\rm MSE$ of $\boldsymbol{\theta}_{\rm RM}$}
			\label{F1_theta_RM}
		\end{minipage}%
	}%
	\subfigure{
		\begin{minipage}[h]{0.24\linewidth}
			\centering
			\includegraphics[width=1.7in]{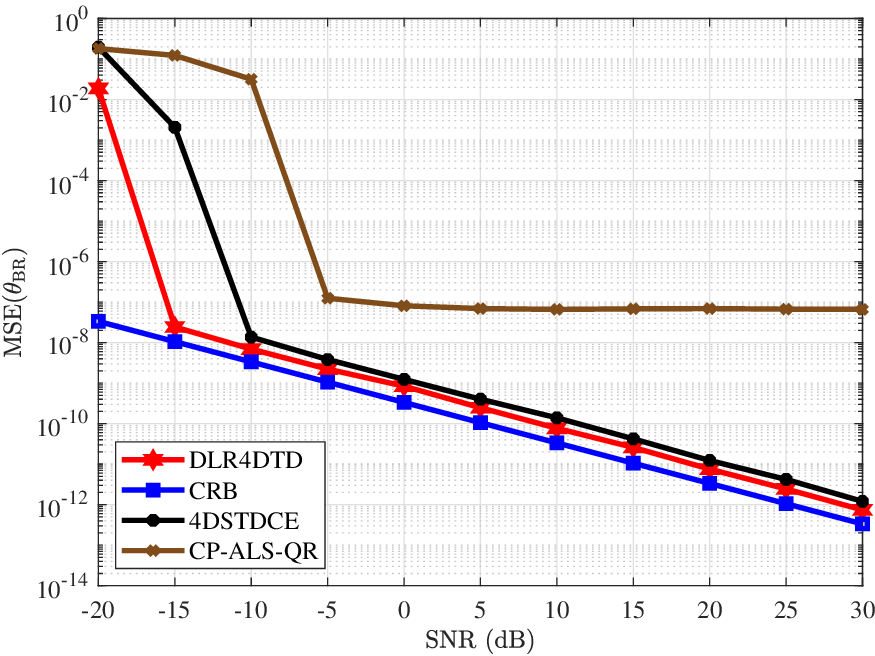}
			\caption*{(c) $\rm MSE$ of $\theta_{\rm BR}$}
			\label{F1_theta_BR}
		\end{minipage}%
	}%
	\subfigure{
		\begin{minipage}[h]{0.24\linewidth}
			\centering
			\includegraphics[width=1.7in]{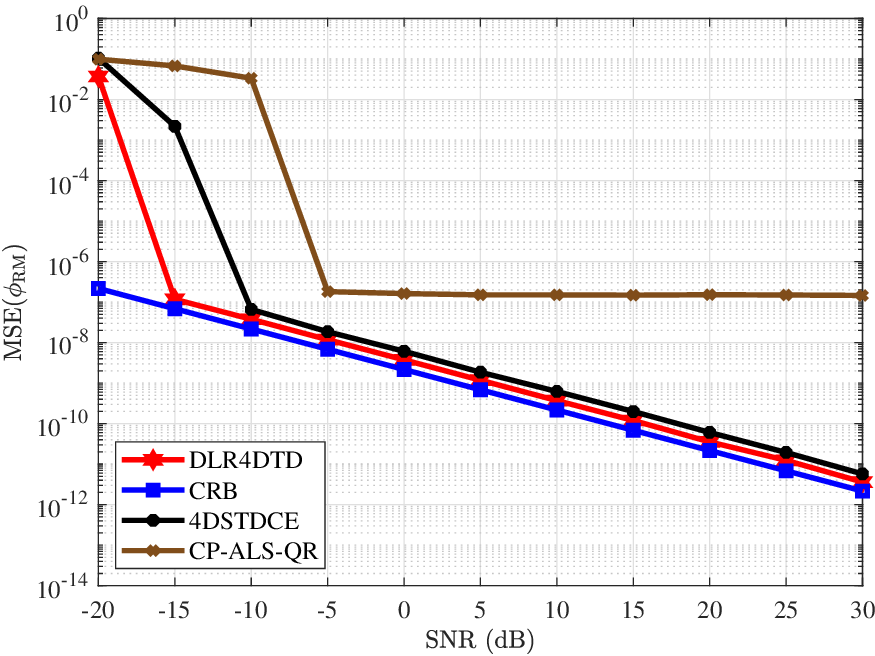}
			\caption*{(d) $\rm MSE$ of $\boldsymbol{\phi}_{\rm RM}$}
			\label{F1_phi_RM}
		\end{minipage}%
	}%
	
	\subfigure{
		\begin{minipage}[h]{0.24\linewidth}
			\centering
			\includegraphics[width=1.7in]{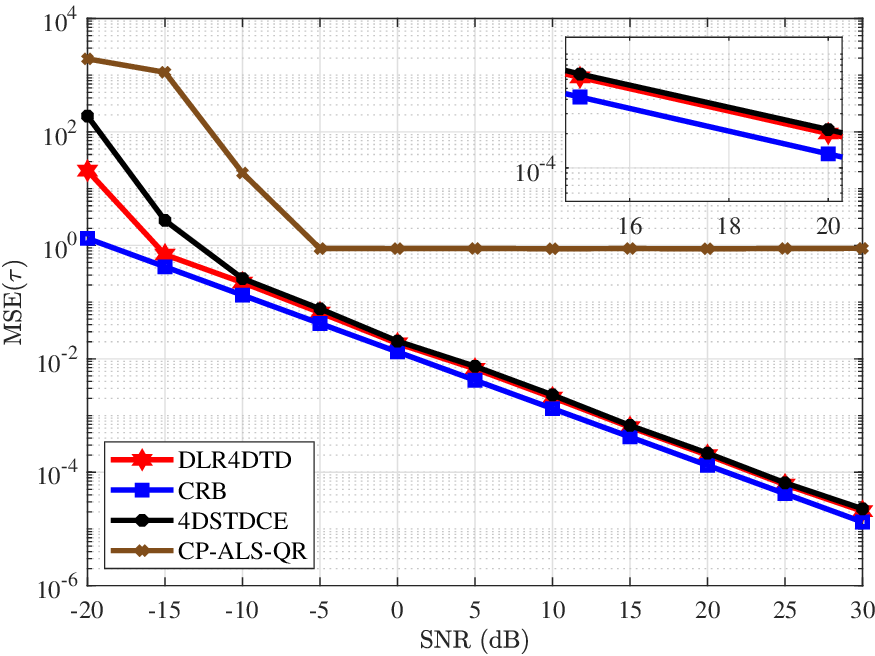}
			\caption*{(e) $\rm MSE$ of $\boldsymbol{\tau}$}
			\label{F1_tau}
		\end{minipage}%
	}%
	\subfigure{
		\begin{minipage}[h]{0.24\linewidth}
			\centering
			\includegraphics[width=1.7in]{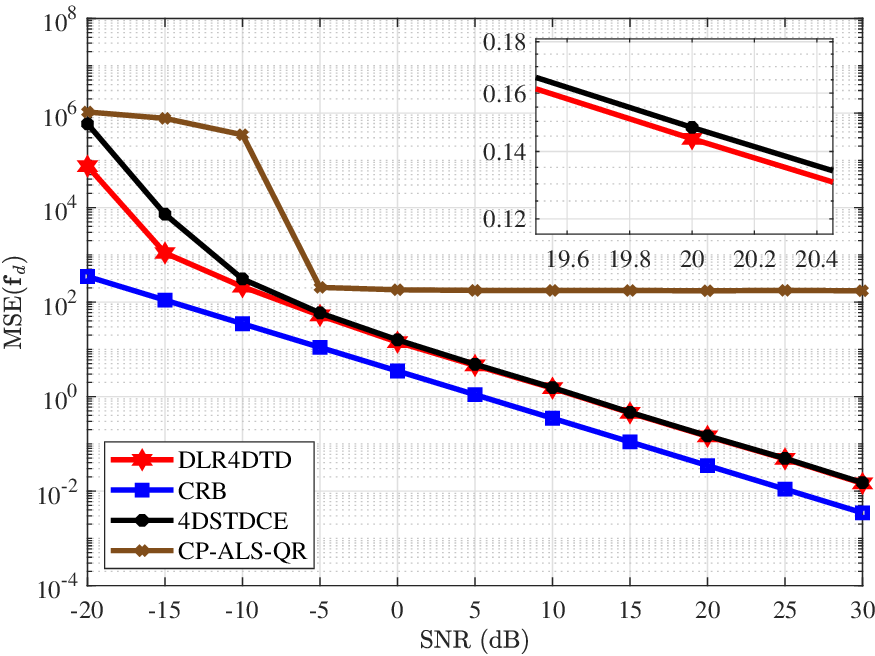}
			\caption*{(f) $\rm MSE$ of ${\bf f}_d$}
			\label{F1_fd}
		\end{minipage}%
	}%
	\subfigure{
		\begin{minipage}[h]{0.24\linewidth}
			\centering
			\includegraphics[width=1.7in]{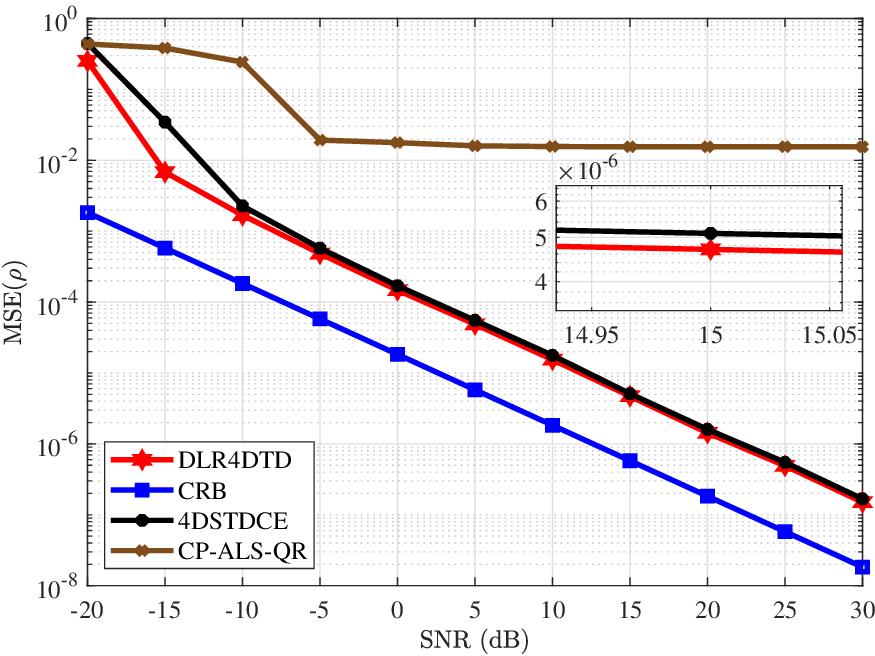}
			\caption*{(g) $\rm MSE$ of $\boldsymbol{\rho}$}
			\label{F1_rho}
		\end{minipage}%
	}%
	\subfigure{
		\begin{minipage}[h]{0.24\linewidth}
			\centering
			\includegraphics[width=1.7in]{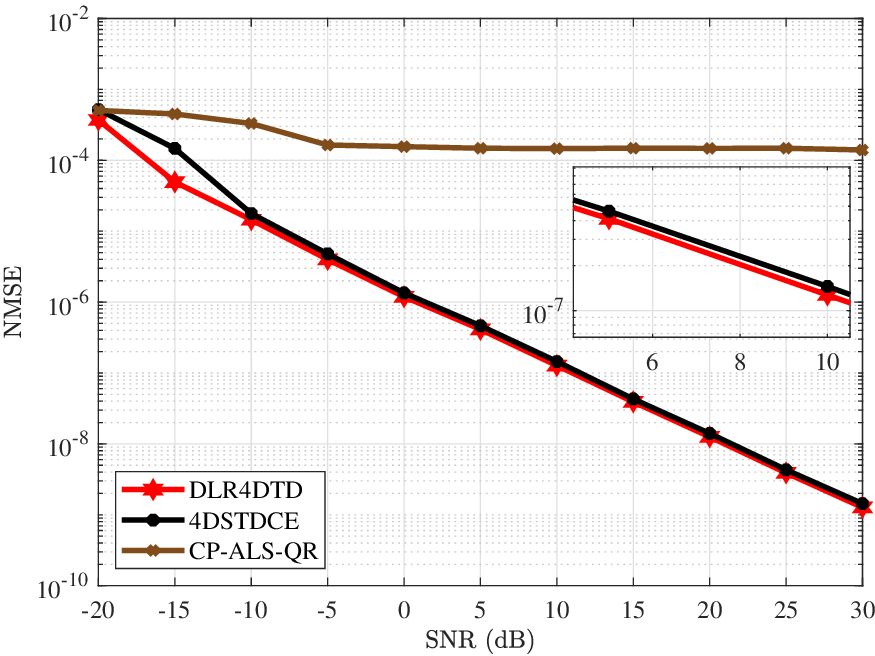}
			\caption*{(h) ${\rm NMSE}$ of cascade channel}
			\label{F1_NMSE}
		\end{minipage}%
	}%
	\centering
	\caption{Comparison of channel estimation performance versus $\rm SNR$.}
	\label{F1}
\end{figure*}
Numerical experiments are conducted to demonstrate the effectiveness of the proposed method. In the simulations, CRB is employed as the benchmark, while the CP-ALS-QR \cite{minster2023cp} and 4DSTDCE are chosen as comparison algorithms for performance evaluation. For the parameter settings, the central carrier frequency and bandwidth of the system are set as $f_c = 30$ GHz and $f_s = 0.12288$ GHz, respectively. Both the BS and MSs are equipped with $N_{\rm BS}=N_{\rm MS}=32$ antennas. The circular RIS with radius $r=20 \lambda$ consists of $N_{\rm R} = 256$ reflecting unit cells. By this design, the distance between RIS array elements can be close to $\lambda / 2$. The number of data streams transmitted simultaneously is set to $N_s = 4$. As for the mmWave channel, there exists one LoS path between the BS and RIS, and $L=4$ paths between the RIS and the MSs. The angular parameters are set as $\theta_{\rm BR} \in (-\pi, -\pi/2)$, $\phi_{\rm BR} \in (0, \pi/2)$, $\theta_{\rm RM}^l \in (0,\pi)$, and $\phi_{\rm RM}^l \in (-\pi /2,0)$, and the channel complex gains $\alpha$ and $\beta_l$ are generated according to a complex normal Gaussian distribution ${\cal CN}(0,1)$. The moving velocity of MS is set as $v = 80$ Km/h. Both the hybrid precoding matrix $\bf F$ and the hybrid combining matrix $\bf W$ are randomly generated from a unit circle, each column of which is normalized to guarantee the power limitation. As for NOMA parameters, the total number of NOMA subcarriers is $N = 256$, among which the first $K = 32$ subcarriers are selected as pilot subcarriers. The subcarrier spacing is set as $\Delta f = 480$ KHz. Besides, the $4$-QAM modulation is employed to modulate the transmitted signal. In the NOMA pair, two MSs are considered, where the power allocation factor of MS $1$ (low-order user) and MS $2$ (high-order user) are set as $\iota_1 = 0.8$ and $\iota_2 = 0.2$, respectively. For the proposed subframe partitioning scheme, there are $M=8$ aggregated slots, each of which has $N_{st} = 8$ half slots, and each half slot has $N_b=7$ NOMA symbols. 
For the proposed DLR4DTD model, parameters $\mu_1$ and $\mu_2$ are chosen by using the cross-validation approach. 

To facilitate the performance analysis, the SNR is defined as ${\rm SNR } = \Vert{\boldsymbol{{\cal Y - N}}^{\bf W}}\Vert_F^2/ \Vert{\boldsymbol{{\cal N}}^{\bf W}}\Vert_F^2$, and the mean square error (MSE) is used to quantify the channel parameters estimation accuracy, defined as
\begin{equation}
\begin{aligned}
{\rm MSE}({\bf x}) = \frac{1}{L} \Vert {\bf x} - { \tilde{\bf x}} \Vert_2^2
\end{aligned}
\label{MSE.eqn}
\end{equation}
Besides, the normalized mean square error (NMSE) is utilized to quantify the cascade channel estimation performance, defined as
\begin{equation}
\begin{aligned}
{\rm NMSE} = \frac{1}{NM} \sum_{m=1}^{M}  \sum_{n=1}^{N} \frac{ \Vert {{\bf H}}^{\bf G}_k[m] - \widetilde{{{\bf H}}^{\bf G}_k}[m] \Vert_F^2}{\Vert {{\bf H}}^{\bf G}_k[m] \Vert_F^2}
\end{aligned}
\label{NMSE.eqn}
\end{equation}
Here, all the simulation results are averaged over independent Monte Carlo trials.

Fig. \ref{F1} shows the channel parameters estimation performance of the discussed methods versus SNR. From Fig. \ref{F1_phi_BR}, it can be observed that the $\rm MSE$ of the estimated channel parameter $\boldsymbol{\theta}_{\rm BR}$ of the discussed methods decreases with increasing $\rm SNR$, in which the proposed method is the closest to CRB. Furthermore, it is worth mentioning that at ${\rm SNR}=-15$ dB, the proposed method still exhibits superior estimation performance, while the performance of 4DSTDCE and CP-ALS-QR methods deteriorates rapidly. This is because the proposed method can efficiently exploit both the global and local low-rank properties of the received signal, whereas the 4DSTDCE method exclusively relies on local low-rank property. This discrepancy renders the 4DSTDCE method particularly vulnerable under low $\rm SNR$, where the high-intensity noise fundamentally compromises the reliability of local low-rank features. The similar $\rm MSE$ results regard to channel parameters $\boldsymbol{\theta}_{\rm RM}$, $\boldsymbol{\theta}_{\rm BR}$, $\boldsymbol{\phi}_{\rm RM}$, $\boldsymbol{\tau}$, ${\bf f}_d$, and $\boldsymbol{\rho}$ can be observed in Figs. \ref{F1_theta_RM}-\ref{F1_rho}. It is worth noting that the CP-ALS-QR method reaches a performance bottleneck at high SNR under these parameters, due to convergence problem \cite{9835123}. To quantify the channel estimation performance, we calculate the $\rm NMSE$ of the cascade channel, as shown in Fig. \ref{F1_NMSE}. The $\rm NMSE$ of the proposed method decreases exponentially with the increase of SNR, while the CP-ALS-QR method hardly decreases at high SNR. Overall, from Fig. \ref{F1}, we can observe that the proposed method achieves the best estimation performance, especially at low $\rm SNR$, and possesses robustness compared with other methods.

\begin{figure}[h]
	\centering
	\includegraphics[width=3 in]{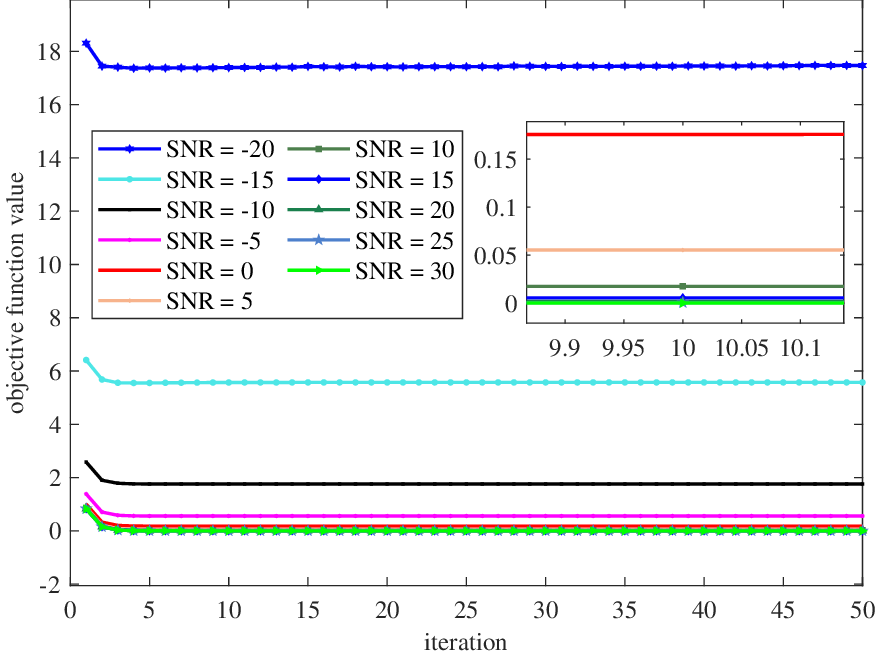}
	\caption{Convergence analysis.}
	\label{convergence}
\end{figure}

Fig. \ref{convergence} shows the objective function values of the proposed method versus iteration times under different $\rm SNR$. It can be observed from Fig. \ref{convergence} that the proposed method can achieve convergence with very few iterations at different $\rm SNR$. It is worth mentioning that the proposed method is not sensitive to initialization, which is one of the main reasons affecting the performance of the CP-ALS-based method.

\begin{figure}[h]
	\centering
	\includegraphics[width=3 in]{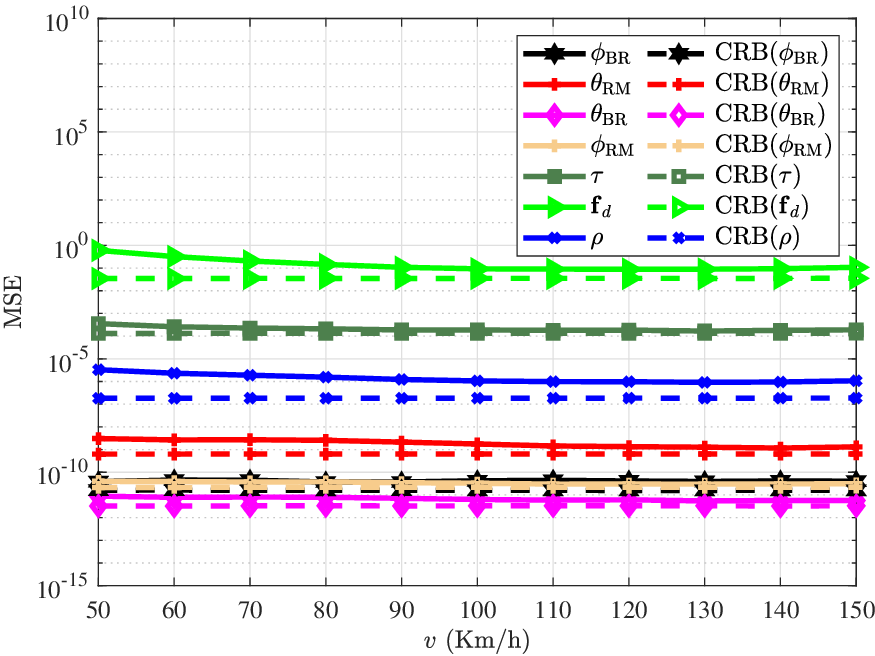}
	\caption{Comparison of estimation performance versus velocity.}
	\label{velocity}
\end{figure}

Furthermore, to verify the performance of the proposed method at different vehicle velocities, we conduct experiments on the channel estimation performance versus velocity when ${\rm SNR} = 20$ dB, whose result is shown in Fig. \ref{velocity}. The results show that the proposed method can achieve stable estimation performance for different vehicle velocities, possessing high adaptability in high-mobility scenarios. 

\section{Conclusion}
In this paper, we proposed a double low-rank 4D tensor decomposition-based channel estimation method for circular RIS-aided mmWave MIMO-NOMA system considering Doppler shift caused by mobility scenarios. By introducing the special structure of circular RIS, the system achieved the unique decoupling of angles, which could not be supported by linear RIS topologies. Firstly, we modeled the received signal as a fourth-order tensor which separated AoA, AoD, time delay, and Doppler shift into four corresponding factor matrices. Subsequently, by leveraging both the global and local low-rank properties of the received signal, we proposed a double low-rank 4D tensor decomposition method, improving the decomposition accuracy, especially at low SNR. Then, a two-stage angles estimation method based on the Jacobi-Angler expansion of the circular RIS was proposed for angle decoupling. Besides, the time delay, Doppler shift, and channel gain could be estimated without ambiguities. Finally, numerical experiments verified the superior channel estimation performance of the proposed method. 

\appendices
\section{Deviation of CRB}	\label{CRB_deviation}
Before calculating the CRB of each parameter, the covariance matrix of the noise vector and the Kronecker product form of ${\bf z}_{(1)}$ are derived for subsequent analysis. Firstly, by performing the mode-$1$ unfolding on noise tensor $\boldsymbol{{\cal N}}^{\bf W} ={\rm Vec}^1_2(\boldsymbol{{\cal N}} \times_1 {\bf W}^T)$, we have ${\bf N}^{\bf W}_{(1)} = {\bf W}^T  {\bf N}_{(1)}$, where ${\bf N}_{(1)}$ is the mode-$1$ unfolding form of $\boldsymbol{{\cal N}}$. Subsequently, the vectorization form of the ${\bf N}^{\bf W}_{(1)}$ can be obtained by
\begin{equation}
\begin{aligned}
{\rm vec}({\bf N}^{\bf W}_{(1)}) &= {\rm vec}({\bf W}^T  {\bf N}_{(1)}) \\
&= ( {\bf I}_{N_b N_{st} K M} \otimes {\bf W}^T) {\rm vec}({\bf N}_{(1)})
\end{aligned}
\label{vec_nw.eqn}
\end{equation}
Since each element of $\boldsymbol{{\cal N}}$ follows an additive i.i.d. zero-mean circularly symmetric complex white Gaussian distribution whose variance is $\sigma^2$, its covariance matrix ${\bf C}_{\boldsymbol{ {\cal N}}}$ and second-order moments ${\bf M}_{\boldsymbol{{\cal N}}}$ can be calculated by

\begin{equation}
\begin{aligned}
{\bf C}_{\boldsymbol{ {\cal N}}} &= {\mathbb E}\{ {\rm vec}({\bf N}^{\bf W}_{(1)}) {\rm vec}({\bf N}^{\bf W}_{(1)})^H \} \\
&=\sigma^2 ({\bf I}_{N_b N_{st} K M} \otimes ({\bf W}^T {\bf W}^*)) \\
{\bf M}_{\boldsymbol{{\cal N}}} &= {\mathbb E}\{{\rm vec}({\bf N}^{\bf W}_{(1)}) {\rm vec}({\bf N}^{\bf W}_{(1)})^T \}  \\
&= {\mathbb E}\{{\rm vec}({\bf N}^{\bf W}_{(1)})^* {\rm vec}({\bf N}^{\bf W}_{(1)})^H \} \\
& = {\bf 0}
\end{aligned}
\label{Cn_calculate.eqn}
\end{equation}

Then, by using the property of Khatri-Rao product, we have 
\begin{equation}
\begin{aligned}
{\bf z}_{(1)} &= {\rm vec}({\bf Z}_{(1)}) =  {\rm vec}({\bf A}({\bf D} \odot {\bf C} \odot {\bf B} )^T)\\
&=({\bf D} \odot {\bf C} \odot {\bf B} \odot {\bf A}) {\bf 1} = \sum_{l=1}^{L} {\bf d}_l \otimes {\bf c}_l \otimes {\bf b}_l \otimes {\bf a}_l
\end{aligned}
\label{p1_calculate.eqn}
\end{equation}

In the following, the detailed complex Fisher information matrix derivation is discussed. The partial derivative of ${\cal L}(\boldsymbol{\eta})$ with respect to ${\phi}_{\rm BR}$ is calculated by 
\begin{equation}
\begin{aligned}
\frac{\partial {\cal L}(\boldsymbol{\eta})}{\partial {\phi}_{\rm BR}}= \left(\frac{\partial {\cal L}(\boldsymbol{\eta})}{\partial {\bf z}_{(1)}}\right)^T  \frac{\partial {\bf z}_{(1)}}{\partial {\phi}_{\rm BR}} +    \left(\frac{\partial {\cal L}(\boldsymbol{\eta})}{\partial {\bf z}_{(1)}^*}\right)^T  \frac{\partial {\bf z}_{(1)}^*}{\partial {\phi}_{\rm BR}} 
\end{aligned}
\label{partial_derivation_phi_BR.eqn}
\end{equation}
where 
\begin{equation}
\begin{aligned}
&\frac{\partial {\cal L}(\boldsymbol{\eta})}{\partial {\bf z}_{(1)}} = {\bf C}_{\boldsymbol{ {\cal N}}}^{-T}({\bf y}_{(1)} - {\bf z}_{(1)})^* = {\bf C}_{\boldsymbol{ {\cal N}}}^{-T} {\rm vec}({\bf N}^{\bf W}_{(1)})^*\\
& \frac{\partial {\cal L}(\boldsymbol{\eta})}{\partial {{\bf z}_{(1)}^*}} =\left (\frac{\partial {\cal L}(\boldsymbol{\eta})}{\partial {\bf z}_{(1)}} \right)^* ~~~~ \frac{\partial {\bf z}_{(1)}^*}{\partial {{\phi}_{\rm BR}}} =\left (\frac{\partial {\bf z}_{(1)}}{\partial  {{\phi}_{\rm BR}}}\right)^* \\
&\frac{\partial {\bf z}_{(1)}}{\partial {{\phi}_{\rm BR}}} =  \sum_{l=1}^{L} {\bf d}_l \otimes {\bf c}_l \otimes {\bf b}_l \otimes \bar{{\bf a}}_l\\
&\bar{{\bf a}}_l= \frac{\partial {\bf a}_l}{\partial {{\phi}_{\rm BR}}} = 
\boldsymbol{\Upsilon}^T{\bf D}_{a1} {\bf a}_{s}(\phi_{\rm BR},\theta_{\rm RM}^l)\\
&{\bf D}_{a1} = -j \pi {\rm sin}(\phi_{\rm BR}) ({\bf I}_{N_{\rm MS}} \otimes {\rm diag}([0,\cdots, {N}_{\rm BS}-1]))
\end{aligned}
\label{partial_derivation_phi_BR_where.eqn}
\end{equation}
Therefore, we have
\begin{equation}
\begin{aligned}
\frac{\partial {\cal L}(\boldsymbol{\eta})}{\partial {\phi}_{\rm BR}} &= {\rm vec}({\bf N}^{\bf W}_{(1)})^H  {\bf C}_{\boldsymbol{ {\cal N}}}^{-1} \frac{\partial {\bf z}_{(1)}}{\partial {\phi}_{\rm BR}} + {\rm vec}({\bf N}^{\bf W}_{(1)})^T  {\bf C}_{\boldsymbol{ {\cal N}}}^{-*} (\frac{\partial {\bf z}_{(1)}}{\partial {{\phi}_{\rm BR}}})^* \\
& = 2 {\rm Re} \left\{  {\rm vec}({\bf N}^{\bf W}_{(1)})^H  {\bf C}_{\boldsymbol{ {\cal N}}}^{-1} \frac{\partial {\bf z}_{(1)}}{\partial {{\phi}_{\rm BR}}}    \right\}
\end{aligned} 
\label{Lp_theta.eqn}
\end{equation}
where ${\rm Re}\{{\cdot}\}$ denotes the real part extraction operation of a complex number. 

Similarly, the partial derivatives with respect to the other channel parameters can be obtained by
\begin{equation}
\begin{aligned}
\frac{\partial {\cal L}(\boldsymbol{\eta})}{\partial {\theta}_{\rm RM}^l} &= 2 {\rm Re} \left\{  {\rm vec}({\bf N}^{\bf W}_{(1)})^H  {\bf C}_{\boldsymbol{ {\cal N}}}^{-1} \frac{\partial {\bf z}_{(1)}}{\partial {{\theta}_{\rm RM}^l}} \right\} \\
\frac{\partial {\cal L}(\boldsymbol{\eta})}{\partial {\theta}_{\rm BR}} &= 2 {\rm Re} \left\{  {\rm vec}({\bf N}^{\bf W}_{(1)})^H  {\bf C}_{\boldsymbol{ {\cal N}}}^{-1} \frac{\partial {\bf z}_{(1)}}{\partial {{\theta}_{\rm BR}}} \right\} \\
\frac{\partial {\cal L}(\boldsymbol{\eta})}{\partial {\phi}_{\rm RM}^l} &= 2 {\rm Re} \left\{  {\rm vec}({\bf N}^{\bf W}_{(1)})^H  {\bf C}_{\boldsymbol{ {\cal N}}}^{-1} \frac{\partial {\bf z}_{(1)}}{\partial {{\phi}_{\rm RM}^l}}  \right\}  \\
\frac{\partial {\cal L}(\boldsymbol{\eta})}{\partial {\tau}_l} &= 2 {\rm Re} \left\{  {\rm vec}({\bf N}^{\bf W}_{(1)})^H  {\bf C}_{\boldsymbol{ {\cal N}}}^{-1} \frac{\partial {\bf z}_{(1)}}{\partial {\tau}_l} \right\} \\
\frac{\partial {\cal L}(\boldsymbol{\eta})}{\partial f_d^l} &= 2 {\rm Re} \left\{  {\rm vec}({\bf N}^{\bf W}_{(1)})^H  {\bf C}_{\boldsymbol{ {\cal N}}}^{-1} \frac{\partial {\bf z}_{(1)}}{\partial f_d^l} \right\}\\
\frac{\partial {\cal L}(\boldsymbol{\eta})}{\partial \rho_l} &=  {\rm vec}({\bf N}^{\bf W}_{(1)})^H  {\bf C}_{\boldsymbol{ {\cal N}}}^{-1} \frac{\partial {\bf z}_{(1)}}{\partial  \rho_l} 
\end{aligned} 
\label{Lp_phi_tau_fl.eqn}
\end{equation}
where
\begin{equation}
\begin{aligned}
&\frac{\partial {\bf z}_{(1)}}{\partial {{\theta}_{\rm RM}^l}}= {\bf d}_l \otimes {\bf c}_l \otimes {\bf b}_l \otimes \tilde{{\bf a}}_l,~\tilde{{\bf a}}_l = 
\boldsymbol{\Upsilon}^T{\bf D}_{a2} {\bf a}_{s}(\phi_{\rm BR},\theta_{\rm RM}^l)\\
&\frac{\partial {\bf z}_{(1)}}{\partial {{\theta}_{\rm BR}}} =  \sum_{l=1}^{L} {\bf d}_l \otimes {\bf c}_l \otimes \bar{{\bf b}}_l \otimes {\bf a}_l,~ \bar{{\bf b}}_l =\boldsymbol{\Xi}^T {\bf D}_{b1}{\bf a}_{r}(\theta_{\rm BR},\phi_{\rm RM}^l) \\
&\frac{\partial {\bf z}_{(1)}}{\partial {{\phi}_{\rm RM}^l}}  =  {\bf d}_l \otimes {\bf c}_l \otimes \tilde{{\bf b}}_l \otimes {\bf a}_l,~ \tilde{{\bf b}}_l = \boldsymbol{\Xi}^T {\bf D}_{b2}{\bf a}_{r}(\theta_{\rm BR},\phi_{\rm RM}^l) \\
& \frac{\partial {\bf z}_{(1)}}{\partial {\tau}_l} = {\bf d}_l \otimes \bar{{\bf c}}_l \otimes {\bf b}_l \otimes {\bf a}_l, ~\bar{{\bf c}}_l = {\bf D}_c{\bf c}_l\\
& \frac{\partial {\bf z}_{(1)}}{\partial f_d^l} =\bar{{\bf d}}_l \otimes {\bf c}_l \otimes {\bf b}_l \otimes {\bf a}_l, ~\bar{{\bf d}}_l = {\bf D}_d{\bf d}_l\\
& \frac{\partial {\bf z}_{(1)}}{\partial  \rho_l} ={\bf d}_l \otimes \tilde{{\bf c}}_l \otimes {\bf b}_l \otimes {\bf a}_l,~\tilde{{\bf c}}_l={\bf g}(\tau_l) \\
&{\bf D}_{a2} =  -j \pi {\rm sin}(\theta_{\rm RM}^l) ({\bf I}_{N_{\rm BS}} \otimes {\rm diag}([0,1,\cdots, {N}_{\rm MS}-1]))\\
& {\bf D}_{b1} = -jk_0r{\rm diag}([ {\rm sin}(\theta_{\rm BR}-\zeta_1),\cdots,{\rm sin}(\theta_{\rm BR}-\zeta_{N_{\rm R}}) ] ) \\
& {\bf D}_{b2} = -jk_0r{\rm diag}([ {\rm sin}({\phi}_{\rm RM}^l-\zeta_1),\cdots,{\rm sin}({\phi}_{\rm RM}^l-\zeta_{N_{\rm R}}) ] )
\end{aligned} 
\label{Lp_phi_tau_fl_partial.eqn}
\end{equation}
\begin{equation*}
\begin{aligned}
& {\bf D}_c = -j \frac{2\pi f_s}{N} {\rm diag}([1,\cdots, K]) \\
& {\bf D}_d = j 2 \pi T_s N_b N_{st} {\rm diag}([1,\cdots, M])
\end{aligned} 
\label{Lp_phi_tau_fl_partial_where.eqn}
\end{equation*}

Then, the entries in the principal minors of ${\cal L}(\boldsymbol{\eta})$ is calculated below. For example, the $(l_1,l_2)$-th entry of ${\mathbb E} \left\{ \left(\frac{\partial {\cal L}(\boldsymbol{\eta})}{\partial \boldsymbol{\theta}_{\rm RM}}\right)^H  \frac{\partial {\cal L}(\boldsymbol{\eta})}{\partial \boldsymbol{\theta}_{\rm RM}} \right\}$ is calculated by 
\begin{equation}
\begin{aligned}
&{\mathbb E} \left\{ \left(\frac{\partial {\cal L}(\boldsymbol{\eta})}{\partial  \theta_{\rm RM}^{l_1}} \right)^*  \frac{\partial {\cal L}(\boldsymbol{\eta})}{\partial  \theta_{\rm RM}^{l_2}} \right\} = \\
&{\mathbb E}   \left\{ \begin{aligned}
({\rm vec}({\bf N}^{\bf W}_{(1)})^H  {\bf C}_{\boldsymbol{ {\cal N}}}^{-1} \frac{\partial {\bf z}_{(1)}}{\partial {{\theta}_{\rm RM}^{l_1}}} +  ({\rm vec}({\bf N}^{\bf W}_{(1)})^H  {\bf C}_{\boldsymbol{ {\cal N}}}^{-1} \frac{\partial {\bf z}_{(1)}}{\partial {{\theta}_{\rm RM}^{l_1}}})^*)^T \\
({\rm vec}({\bf N}^{\bf W}_{(1)})^H  {\bf C}_{\boldsymbol{ {\cal N}}}^{-1} \frac{\partial {\bf z}_{(1)}}{\partial {{\theta}_{\rm RM}^{l_2}}} +  ({\rm vec}({\bf N}^{\bf W}_{(1)})^H  {\bf C}_{\boldsymbol{ {\cal N}}}^{-1} \frac{\partial {\bf z}_{(1)}}{\partial {{\theta}_{\rm RM}^{l_2}}})^*)
\end{aligned}
\right \} \\
&=\left( \frac{\partial {\bf z}_{(1)}}{\partial {{\theta}_{\rm RM}^{l_1}}}\right)^T {\bf C}_{\boldsymbol{ {\cal N}}}^{-T} \left( \frac{\partial {\bf z}_{(1)}}{\partial {{\theta}_{\rm RM}^{l_2}}}\right)^*+\left( \frac{\partial {\bf z}_{(1)}}{\partial {{\theta}_{\rm RM}^{l_1}}}\right)^H {\bf C}_{\boldsymbol{ {\cal N}}}^{-1}  \frac{\partial {\bf z}_{(1)}}{\partial {{\theta}_{\rm RM}^{l_2}}} \\
& = 2{\rm Re} \left\{     \left( \frac{\partial {\bf z}_{(1)}}{\partial {{\theta}_{\rm RM}^{l_1}}}\right)^H {\bf C}_{\boldsymbol{ {\cal N}}}^{-1}  \frac{\partial {\bf z}_{(1)}}{\partial {{\theta}_{\rm RM}^{l_2}}}     \right\}
\end{aligned}
\label{partial_derivation_thetal1l2.eqn}
\end{equation}

And the elements in the off-principal minors of ${\cal L}(\boldsymbol{\eta})$, such as the $(l_1, l_2)$-th entry of ${\mathbb E} \left\{ \left(\frac{\partial {\cal L}(\boldsymbol{\eta})}{\partial \boldsymbol{\theta}_{\rm RM}}\right)^H  \frac{\partial {\cal L}(\boldsymbol{\eta})}{\partial \boldsymbol{\phi}_{\rm RM}} \right\}$, can be calculated by 
\begin{equation}
\begin{aligned}
&{\mathbb E} \left\{ \left(\frac{\partial {\cal L}(\boldsymbol{\eta})}{\partial  \theta_{\rm RM}^{l_1}} \right)^*  \frac{\partial {\cal L}(\boldsymbol{\eta})}{\partial  \phi_{\rm RM}^{l_2}} \right\} = 2{\rm Re} \left\{     \left( \frac{\partial {\bf z}_{(1)}}{\partial {{\theta}_{\rm RM}^{l_1}}}\right)^H {\bf C}_{\boldsymbol{ {\cal N}}}^{-1}  \frac{\partial {\bf z}_{(1)}}{\partial {{\phi}_{\rm RM}^{l_2}}}     \right\}
\end{aligned}
\label{partial_derivation_thetal1phil2.eqn}
\end{equation}

Similarly, we can arrive at
\begin{equation}
\begin{aligned}
&{\mathbb E} \left\{ \left(\frac{\partial {\cal L}(\boldsymbol{\eta})}{\partial  {\kappa}_{l_1}} \right)^*  \frac{\partial {\cal L}(\boldsymbol{\eta})}{\partial  {\upsilon}_{l_2}} \right\}  = 2{\rm Re} \left\{     \left( \frac{\partial {\bf z}_{(1)}}{\partial {\kappa_{l_1}}}\right)^H {\bf C}_{\boldsymbol{ {\cal N}}}^{-1}  \frac{\partial {\bf z}_{(1)}}{\partial {\upsilon_{l_2}}}     \right\} \\
&{\mathbb E} \left\{ \left(\frac{\partial {\cal L}(\boldsymbol{\eta})}{\partial {\rho}_{l_1}} \right)^*  \frac{\partial {\cal L}(\boldsymbol{\eta})}{\partial  {\kappa}_{l_2}} \right\}  =     \left( \frac{\partial {\bf z}_{(1)}}{\partial {{\rho}_{l_1}}}\right)^H {\bf C}_{\boldsymbol{ {\cal N}}}^{-1}  \frac{\partial {\bf z}_{(1)}}{\partial { {\kappa}_{l_2}}}   \\
&{\mathbb E} \left\{ \left(\frac{\partial {\cal L}(\boldsymbol{\eta})}{\partial \kappa_{l_1}} \right)^*  \frac{\partial {\cal L}(\boldsymbol{\eta})}{\partial  \rho_{l_2}} \right\}  = \left( \frac{\partial {\bf z}_{(1)}}{\partial {\kappa_{l_1}}}\right)^H {\bf C}_{\boldsymbol{ {\cal N}}}^{-1}  \frac{\partial {\bf z}_{(1)}}{\partial {\rho_{l_2}}} 
\end{aligned}
\label{partial_derivation_thetaphitau.eqn}
\end{equation}
where $\kappa$ and $\upsilon$ represent the any one of the set $\{{\phi}_{\rm BR},{\theta}_{\rm RM}, {\theta}_{\rm BR}, {\phi}_{\rm RM}, {\tau_l}, \tilde{f}_d, {\rho} \}$. 

Therefore, we obtain the FIM, so that the CRB for $\boldsymbol{\eta}$ can be calculated from \eqref{CRB.eqn}.

\vspace{-0.2em}

\vfill

\end{document}